\documentclass[a4paper]{spie} 
\usepackage[dvipdfmx]{graphicx}
\usepackage{amssymb}
\usepackage{mediabb}
\usepackage{color}

\newcommand{\degree}{~$^{\circ}\mbox{C}$}
\newcommand{\chandra}{Chandra}
\newcommand{\xmm}{XMM-Newton}
\newcommand{\suzaku}{Suzaku}
\newcommand{\integral}{INTEGRAL}

\newcommand{\nustar}{NuSTAR}
\newcommand{\astroh}{ASTRO-H}

\newcommand{\red}{\textcolor{black}}
\newcommand{\green}{\textcolor{black}}
\newcommand{\blue}{\textcolor{black}}

\title{The ASTRO-H X-ray Astronomy Satellite} 

\normalsize
 \normalsize
\author{
\mbox{\blue{Tadayuki~Takahashi}\supit{a},}
\mbox{\blue{Kazuhisa~Mitsuda}\supit{a},}
\mbox{\blue{Richard~Kelley}\supit{b},}
\mbox{\blue{Felix~Aharonian}\supit{c},}
\mbox{\blue{Hiroki~Akamatsu}\supit{d},}
\mbox{\blue{Fumie~Akimoto}\supit{e},}
\mbox{\blue{Steve~Allen}\supit{f},}
\mbox{\blue{Naohisa~Anabuki}\supit{g},}
\mbox{\blue{Lorella~Angelini}\supit{b},}
\mbox{\blue{Keith~Arnaud}\supit{h},}
\mbox{\green{Makoto~Asai}\supit{f},}
\mbox{\blue{Marc~Audard}\supit{i},}
\mbox{\blue{Hisamitsu~Awaki}\supit{j},}
\mbox{\green{Philipp~Azzarello}\supit{i},}
\mbox{\green{Chris~Baluta}\supit{a},}
\mbox{\blue{Aya~Bamba}\supit{k},}
\mbox{\green{Nobutaka~Bando}\supit{a},}
\mbox{\blue{Marshall~Bautz}\supit{l},}
\mbox{\green{Thomas~Bialas}\supit{b},} 
\mbox{\blue{Roger~Blandford}\supit{f},}
\mbox{\green{Kevin~Boyce}\supit{b},}
\mbox{\blue{Laura~Brenneman}\supit{b},} 
\mbox{\blue{Greg~Brown}\supit{m},}
\mbox{\blue{Edward~Cackett}\supit{n},}
\mbox{\green{Edgar~Canavan}\supit{b},} 
\mbox{\blue{Maria~Chernyakova}\supit{c},}
\mbox{\blue{Meng~Chiao}\supit{b},} 
\mbox{\blue{Paolo~Coppi}\supit{o},}
\mbox{\blue{Elisa~Costantini}\supit{d},}
\mbox{\blue{Jelle~de Plaa}\supit{d},}
\mbox{\blue{Jan-Willem~den Herder}\supit{d},}
\mbox{\green{Michael~DiPirro}\supit{b},}
\mbox{\blue{Chris~Done}\supit{p},}
\mbox{\blue{Tadayasu~Dotani}\supit{a},}
\mbox{\green{John~Doty}\supit{q},}
\mbox{\blue{Ken~Ebisawa}\supit{a},}
\mbox{\blue{Megan~Eckart}\supit{b},}
\mbox{\blue{Teruaki~Enoto}\supit{r},}
\mbox{\blue{Yuichiro~Ezoe}\supit{s},}
\mbox{\blue{Andrew~Fabian}\supit{n},}
\mbox{\blue{Carlo~Ferrigno}\supit{i},}
\mbox{\blue{Adam~Foster}\supit{t},}
\mbox{\blue{Ryuichi~Fujimoto}\supit{u},}
\mbox{\blue{Yasushi~Fukazawa}\supit{v},}
\mbox{\blue{Stefan~Funk}\supit{f},}
\mbox{\blue{Akihiro~Furuzawa}\supit{e},}
\mbox{\blue{Massimiliano~Galeazzi}\supit{w},}
\mbox{\blue{Luigi~Gallo}\supit{x},}
\mbox{\blue{Poshak~Gandhi}\supit{p},}
\mbox{\green{Kirk~Gilmore}\supit{f},}
\mbox{\blue{Matteo~Guainazzi}\supit{y},} 
\mbox{\green{Daniel~Haas}\supit{d},}
\mbox{\blue{Yoshito~Haba}\supit{z},}
\mbox{\blue{Kenji~Hamaguchi}\supit{h},}
\mbox{\green{Atsushi~Harayama}\supit{a},} 
\mbox{\blue{Isamu~Hatsukade}\supit{aa},}
\mbox{\blue{Takayuki~Hayashi}\supit{a},}
\mbox{\blue{Katsuhiro~Hayashi}\supit{a},} 
\mbox{\blue{Kiyoshi~Hayashida}\supit{g},}
\mbox{\blue{Junko~Hiraga}\supit{ab},}
\mbox{\green{Kazuyuki~Hirose}\supit{a},}
\mbox{\blue{Ann~Hornschemeier}\supit{b},}
\mbox{\blue{Akio~Hoshino}\supit{ac},}
\mbox{\blue{John~Hughes}\supit{ad},}
\mbox{\blue{Una~Hwang}\supit{ae},}
\mbox{\blue{Ryo~Iizuka}\supit{a},}
\mbox{\blue{Yoshiyuki~Inoue}\supit{a},}
\mbox{\blue{Kazunori~Ishibashi}\supit{e},}
\mbox{\blue{Manabu~Ishida}\supit{a},}
\mbox{\blue{Kumi~Ishikawa}\supit{r},} 
\mbox{\green{Kosei~Ishimura}\supit{a},}
\mbox{\blue{Yoshitaka~Ishisaki}\supit{s},}
\mbox{\blue{Masayuki~Ito}\supit{af},}
\mbox{\green{Naoko~Iwata}\supit{a},}
\mbox{\blue{Naoko~Iyomoto}\supit{ag},}
\mbox{\green{Chris~Jewell}\supit{ah},} 
\mbox{\blue{Jelle~Kaastra}\supit{d},}
\mbox{\blue{Timothy~Kallman}\supit{b},}
\mbox{\blue{Tuneyoshi~Kamae}\supit{f},}
\mbox{\blue{Jun~Kataoka}\supit{ai},}
\mbox{\blue{Satoru~Katsuda}\supit{a},}
\mbox{\blue{Junichiro~Katsuta}\supit{v},} 
\mbox{\blue{Madoka~Kawaharada}\supit{a},}
\mbox{\blue{Nobuyuki~Kawai}\supit{aj},}
\mbox{\green{Taro~Kawano}\supit{a},} 
\mbox{\green{Shigeo~Kawasaki}\supit{a},}
\mbox{\blue{Dmitry~Khangulyan}\supit{a},}
\mbox{\blue{Caroline~Kilbourne}\supit{b},}
\mbox{\green{Mark~Kimball}\supit{b},} 
\mbox{\blue{Masashi~Kimura}\supit{ak},}
\mbox{\blue{Shunji~Kitamoto}\supit{ac},}
\mbox{\blue{Tetsu~Kitayama}\supit{al},}
\mbox{\blue{Takayoshi~Kohmura}\supit{am},}
\mbox{\blue{Motohide~Kokubun}\supit{a},}
\mbox{\blue{Saori~Konami}\supit{s},} 
\mbox{\green{Tatsuro~Kosaka}\supit{an},}
\mbox{\green{Alex~Koujelev}\supit{ao},}
\mbox{\blue{Katsuji~Koyama}\supit{ap},}
\mbox{\blue{Hans~Krimm}\supit{b},}
\mbox{\blue{Aya~Kubota}\supit{aq},}
\mbox{\blue{Hideyo~Kunieda}\supit{e},}
\mbox{\blue{Stephanie~LaMassa}\supit{o},}
\mbox{\blue{Philippe~Laurent}\supit{ar},}
\mbox{\blue{Fran\c{c}ois~Lebrun}\supit{ar},}
\mbox{\blue{Maurice~Leutenegger}\supit{b},}
\mbox{\blue{Olivier~Limousin}\supit{ar},}
\mbox{\blue{Michael~Loewenstein}\supit{b},}
\mbox{\blue{Knox~Long}\supit{as},}
\mbox{\blue{David~Lumb}\supit{ah},}
\mbox{\blue{Grzegorz~Madejski}\supit{f},}
\mbox{\blue{Yoshitomo~Maeda}\supit{a},}
\mbox{\blue{Kazuo~Makishima}\supit{ab},}
\mbox{\blue{Maxim~Markevitch}\supit{b},}
\mbox{\green{Candace~Masters}\supit{b},} 
\mbox{\blue{Hironori~Matsumoto}\supit{e},}
\mbox{\blue{Kyoko~Matsushita}\supit{at},}
\mbox{\blue{Dan~McCammon}\supit{au},}
\mbox{\green{Daniel~Mcguinness}\supit{b},} 
\mbox{\blue{Brian~McNamara}\supit{av},}
\mbox{\green{Joseph~Miko}\supit{b},} 
\mbox{\blue{Jon~Miller}\supit{aw},}
\mbox{\blue{Eric~Miller}\supit{l},}
\mbox{\blue{Shin~Mineshige}\supit{ax},}
\mbox{\green{Kenji~Minesugi}\supit{a},}
\mbox{\blue{Ikuyuki~Mitsuishi}\supit{e},}
\mbox{\blue{Takuya~Miyazawa}\supit{e},}
\mbox{\blue{Tsunefumi~Mizuno}\supit{v},}
\mbox{\blue{Koji~Mori}\supit{aa},}
\mbox{\blue{Hideyuki~Mori}\supit{e},}
\mbox{\green{Franco~Moroso}\supit{ao},} 
\mbox{\green{Theodore~Muench}\supit{b},} 
\mbox{\blue{Koji~Mukai}\supit{b},}
\mbox{\blue{Hiroshi~Murakami}\supit{ay},}
\mbox{\blue{Toshio~Murakami}\supit{u},}
\mbox{\blue{Richard~Mushotzky}\supit{h},}
\mbox{\green{Housei~Nagano}\supit{e},}
\mbox{\blue{Ryo~Nagino}\supit{g},}
\mbox{\blue{Takao~Nakagawa}\supit{a},}
\mbox{\blue{Hiroshi~Nakajima}\supit{g},}
\mbox{\blue{Takeshi~Nakamori}\supit{az},}
\mbox{\blue{Shinya~Nakashima}\supit{a},} 
\mbox{\blue{Kazuhiro~Nakazawa}\supit{ab},}
\mbox{\green{Yoshiharu~Namba}\supit{ba},}
\mbox{\green{Chikara~Natsukari}\supit{a},}
\mbox{\green{Yusuke~Nishioka}\supit{aa},}
\mbox{\blue{Masayoshi~Nobukawa}\supit{ap},}
\mbox{\blue{Hirofumi~Noda}\supit{r},} 
\mbox{\blue{Masaharu~Nomachi}\supit{bb},}
\mbox{\blue{Steve~O' Dell}\supit{bc},}
\mbox{\blue{Hirokazu~Odaka}\supit{a},}
\mbox{\green{Hiroyuki~Ogawa}\supit{a},}
\mbox{\green{Mina~Ogawa}\supit{a},}
\mbox{\green{Keiji~Ogi}\supit{j},}
\mbox{\blue{Takaya~Ohashi}\supit{s},}
\mbox{\blue{Masanori~Ohno}\supit{v},}
\mbox{\green{Masayuki~Ohta}\supit{a},}
\mbox{\blue{Takashi~Okajima}\supit{b},}
\mbox{\green{Atsushi~Okamoto}\supit{ak},}
\mbox{\green{Tsuyoshi~Okazaki}\supit{a},}
\mbox{\blue{Naomi~Ota}\supit{bd},}
\mbox{\blue{Masanobu~Ozaki}\supit{a},}
\mbox{\blue{Frits~Paerels}\supit{be},}
\mbox{\blue{St\'{e}phane~Paltani}\supit{i},}
\mbox{\blue{Arvind~Parmar}\supit{y},}
\mbox{\blue{Robert~Petre}\supit{b},}
\mbox{\blue{Ciro~Pinto}\supit{n},} 
\mbox{\blue{Martin~Pohl}\supit{i},}
\mbox{\green{James~Pontius}\supit{b},} 
\mbox{\blue{F. Scott~Porter}\supit{b},}
\mbox{\blue{Katja~Pottschmidt}\supit{b},} 
\mbox{\blue{Brian~Ramsey}\supit{bc},}
\mbox{\blue{Rubens~Reis}\supit{aw},}
\mbox{\blue{Christopher~Reynolds}\supit{h},}
\mbox{\blue{Claudio~Ricci}\supit{ax},} 
\mbox{\blue{Helen~Russell}\supit{n},}
\mbox{\blue{Samar~Safi-Harb}\supit{bf},}
\mbox{\blue{Shinya~Saito}\supit{a},} 
\mbox{\green{Shin-ichiro~Sakai}\supit{a},}
\mbox{\blue{Hiroaki~Sameshima}\supit{a},}
\mbox{\blue{Goro~Sato}\supit{ai},}
\mbox{\green{Yoichi~Sato}\supit{ak},}
\mbox{\blue{Kosuke~Sato}\supit{at},}
\mbox{\blue{Rie~Sato}\supit{a},}
\mbox{\blue{Makoto~Sawada}\supit{k},}
\mbox{\blue{Peter~Serlemitsos}\supit{b},}
\mbox{\blue{Hiromi~Seta}\supit{bg},}
\mbox{\green{Yasuko~Shibano}\supit{a},}
\mbox{\green{Maki~Shida}\supit{a},}
\mbox{\green{Takanobu~Shimada}\supit{a},}
\mbox{\green{Keisuke~Shinozaki}\supit{ak},}
\mbox{\green{Peter~Shirron}\supit{b},}
\mbox{\blue{Aurora~Simionescu}\supit{a},}
\mbox{\green{Cynthia~Simmons}\supit{b},}
\mbox{\blue{Randall~Smith}\supit{t},}
\mbox{\green{Gary~Sneiderman}\supit{b},}
\mbox{\blue{Yang~Soong}\supit{b},}
\mbox{\blue{{\L}ukasz~Stawarz}\supit{a},}
\mbox{\blue{Yasuharu~Sugawara}\supit{bh},}
\mbox{\green{Hiroyuki~Sugita}\supit{ak},}
\mbox{\blue{Satoshi~Sugita}\supit{j},}
\mbox{\blue{Andrew~Szymkowiak}\supit{o},}
\mbox{\blue{Hiroyasu~Tajima}\supit{e},}
\mbox{\blue{Hiromitsu~Takahashi}\supit{v},}
\mbox{\blue{Hiroaki~Takahashi}\supit{g},} 
\mbox{\green{Shin-ichiro~Takeda}\supit{a},}
\mbox{\blue{Yoh~Takei}\supit{a},}
\mbox{\blue{Toru~Tamagawa}\supit{r},}
\mbox{\blue{Takayuki~Tamura}\supit{a},}
\mbox{\blue{Keisuke~Tamura}\supit{e},}
\mbox{\blue{Takaaki~Tanaka}\supit{ap},}
\mbox{\blue{Yasuo~Tanaka}\supit{a},}
\mbox{\blue{Yasuyuki~Tanaka}\supit{v},} 
\mbox{\blue{Makoto~Tashiro}\supit{bg},}
\mbox{\blue{Yuzuru~Tawara}\supit{e},}
\mbox{\blue{Yukikatsu~Terada}\supit{bg},}
\mbox{\blue{Yuichi~Terashima}\supit{j},}
\mbox{\blue{Francesco~Tombesi}\supit{b},}
\mbox{\blue{Hiroshi~Tomida}\supit{ak},}
\mbox{\blue{Yohko~Tsuboi}\supit{bh},}
\mbox{\blue{Masahiro~Tsujimoto}\supit{a},}
\mbox{\blue{Hiroshi~Tsunemi}\supit{g},}
\mbox{\blue{Takeshi~Tsuru}\supit{ap},}
\mbox{\blue{Hiroyuki~Uchida}\supit{ap},}
\mbox{\blue{Yasunobu~Uchiyama}\supit{ac},}
\mbox{\blue{Hideki~Uchiyama}\supit{bi},}
\mbox{\blue{Yoshihiro~Ueda}\supit{ax},}
\mbox{\blue{Shutaro~Ueda}\supit{g},} 
\mbox{\blue{Shiro~Ueno}\supit{ak},}
\mbox{\blue{Shinichiro~Uno}\supit{bj},}
\mbox{\blue{Meg~Urry}\supit{o},}
\mbox{\blue{Eugenio~Ursino}\supit{w},}
\mbox{\blue{Cor de~Vries}\supit{d},}
\mbox{\green{Atsushi~Wada}\supit{a},}
\mbox{\blue{Shin~Watanabe}\supit{a},}
\mbox{\green{Tomomi~Watanabe}\supit{b},} 
\mbox{\blue{Norbert~Werner}\supit{f},}
\mbox{\blue{Nicholas~White}\supit{b},}
\mbox{\blue{Dan~Wilkins}\supit{x},} 
\mbox{\green{Takahiro~Yamada}\supit{a},}
\mbox{\blue{Shinya~Yamada}\supit{s},}
\mbox{\blue{Hiroya~Yamaguchi}\supit{b},}
\mbox{\blue{Kazutaka~Yamaoka}\supit{e},} 
\mbox{\blue{Noriko~Yamasaki}\supit{a},}
\mbox{\blue{Makoto~Yamauchi}\supit{aa},}
\mbox{\blue{Shigeo~Yamauchi}\supit{bd},}
\mbox{\blue{Tahir~Yaqoob}\supit{b},} 
\mbox{\blue{Yoichi~Yatsu}\supit{aj},}
\mbox{\blue{Daisuke~Yonetoku}\supit{u},}
\mbox{\blue{Atsumasa~Yoshida}\supit{k},}
\mbox{\blue{Takayuki~Yuasa}\supit{r},}
\mbox{\blue{Irina~Zhuravleva}\supit{f},} 
\mbox{\blue{Abderahmen~Zoghbi}\supit{h},} 
\mbox{\blue{John~ZuHone}\supit{b},} 
\skiplinehalf
\supit{a}Institute of Space and Astronautical Science (ISAS), Japan Aerospace Exploration Agency (JAXA), Kanagawa 252-5210, Japan;
\supit{b}NASA/Goddard Space Flight Center, MD 20771, USA;
\supit{c}Astronomy and Astrophysics Section, Dublin Institute for Advanced Studies, Dublin 2, Ireland;
\supit{d}SRON Netherlands Institute for Space Research, Utrecht, The Netherlands;
\supit{e}Department of Physics, Nagoya University, Aichi 338-8570, Japan;
\supit{f}Kavli Institute for Particle Astrophysics and Cosmology, Stanford University, CA 94305, USA;
\supit{g}Department of Earth and Space Science, Osaka University, Osaka 560-0043, Japan;
\supit{h}Department of Astronomy, University of Maryland, MD 20742, USA;
\supit{i}Universit\'{e} de Gen\`{e}ve, Gen\`{e}ve 4, Switzerland;
\supit{j}Department of Physics, Ehime University, Ehime 790-8577, Japan;
\supit{k}Department of Physics and Mathematics, Aoyama Gakuin University, Kanagawa 229-8558, Japan;
\supit{l}Kavli Institute for Astrophysics and Space Research, Massachusetts Institute of Technology, MA 02139, USA;
\supit{m}Lawrence Livermore National Laboratory, CA 94550, USA;
\supit{n}Institute of Astronomy, Cambridge University, CB3 0HA, UK;
\supit{o}Yale Center for Astronomy and Astrophysics, Yale University, CT 06520-8121, USA;
\supit{p}Department of Physics, University of Durham, DH1 3LE, UK;
\supit{q}Noqsi Aerospace Ltd., CO 80470, USA;
\supit{r}RIKEN, Saitama 351-0198, Japan;
\supit{s}Department of Physics, Tokyo Metropolitan University, Tokyo 192-0397, Japan;
\supit{t}Harvard-Smithsonian Center for Astrophysics, MA 02138, USA;
\supit{u}Faculty of Mathematics and Physics, Kanazawa University, Ishikawa 920-1192, Japan;
\supit{v}Department of Physical Science, Hiroshima University, Hiroshima 739-8526, Japan;
\supit{w}Physics Department, University of Miami, FL 33124, USA;
\supit{x}Department of Astronomy and Physics, Saint Mary's University, Nova Scotia B3H 3C3, Canada;
\supit{y}European Space Agency (ESA), European Space Astronomy Centre (ESAC), Madrid, Spain;
\supit{z}Department of Physics and Astronomy, Aichi University of Education, Aichi 448-8543, Japan;
\supit{aa}Department of Applied Physics, University of Miyazaki, Miyazaki 889-2192, Japan;
\supit{ab}Department of Physics, University of Tokyo, Tokyo 113-0033, Japan;
\supit{ac}Department of Physics, Rikkyo University, Tokyo 171-8501, Japan;
\supit{ad}Department of Physics and Astronomy, Rutgers University, NJ 08854-8019, USA;
\supit{ae}Department of Physics and Astronomy, Johns Hopkins University, MD 21218, USA;
\supit{af}Faculty of Human Development, Kobe University, Hyogo 657-8501, Japan;
\supit{ag}Kyushu University, Fukuoka 819-0395, Japan;
\supit{ah}European Space Agency (ESA), European Space Research and Technology Centre (ESTEC), 2200 AG Noordwijk, The Netherlands;
\supit{ai}Research Institute for Science and Engineering, Waseda University, Tokyo 169-8555, Japan;
\supit{aj}Department of Physics, Tokyo Institute of Technology, Tokyo 152-8551, Japan;
\supit{ak}Tsukuba Space Center (TKSC), Japan Aerospace Exploration Agency (JAXA), Ibaraki 305-8505, Japan;
\supit{al}Department of Physics, Toho University, Chiba 274-8510, Japan;
\supit{am}Department of Physics, Tokyo University of Science, Chiba 278-8510, Japan;
\supit{an}School of Systems Engineering, Kochi University of Technology, Kochi 782-8502, Japan;
\supit{ao}Space Exploration Development Space Exploration, Canadian Space Agency John H. Chapman Space Centre, QC J3Y 8Y9, Canada;
\supit{ap}Department of Physics, Kyoto University, Kyoto 606-8502, Japan;
\supit{aq}Department of Electronic Information Systems, Shibaura Institute of Technology, Saitama 337-8570, Japan;
\supit{ar}IRFU/Service d'Astrophysique, CEA Saclay, 91191 Gif-sur-Yvette Cedex, France;
\supit{as}Space Telescope Science Institute, MD 21218, USA;
\supit{at}Department of Physics, Tokyo University of Science, Tokyo 162-8601, Japan;
\supit{au}Department of Physics, University of Wisconsin, WI 53706, USA;
\supit{av}University of Waterloo, Ontario N2L 3G1, Canada;
\supit{aw}Department of Astronomy, University of Michigan, MI 48109, USA;
\supit{ax}Department of Astronomy, Kyoto University, Kyoto 606-8502, Japan;
\supit{ay}Department of Information Science, Faculty of Liberal Arts, Tohoku Gakuin University, Miyagi 981-3193, Japan;
\supit{az}Department of Physics, Faculty of Science, Yamagata University, Yamagata 990-8560, Japan;
\supit{ba}Department of Mechanical Engineering, Chubu University, Aichi 487-8501, Japan;
\supit{bb}Laboratory of Nuclear Studies, Osaka University, Osaka 560-0043, Japan;
\supit{bc}NASA/Marshall Space Flight Center, AL 35812, USA;
\supit{bd}Department of Physics, Faculty of Science, Nara Women's University, Nara 630-8506, Japan;
\supit{be}Department of Astronomy, Columbia University, NY 10027, USA;
\supit{bf}Department of Physics and Astronomy, University of Manitoba, MB R3T 2N2, Canada;
\supit{bg}Department of Physics, Saitama University, Saitama 338-8570, Japan;
\supit{bh}Department of Physics, Chuo University, Tokyo 112-8551, Japan;
\supit{bi}Science Education, Faculty of Education, Shizuoka University, Shizuoka 422-8529, Japan;
\supit{bj}Faculty of Social and Information Sciences, Nihon Fukushi University, Aichi 475-0012, Japan;
}
\authorinfo{ASTRO-H Web site: http://astro-h.isas.jaxa.jp/en/}

\begin{document}

  \maketitle 

\begin{abstract}

The joint JAXA/NASA ASTRO-H mission is the sixth in a series of highly successful X-ray missions developed by the Institute of Space and Astronautical Science (ISAS), with a planned launch in 2015.
The ASTRO-H mission is equipped with a suite of sensitive instruments with the highest energy resolution ever achieved at E $>$ 3 keV and a wide energy range spanning four decades in energy from soft X-rays to gamma-rays. The simultaneous broad band pass, coupled with the high spectral resolution of $\Delta E $ $\leqq$ 7~eV of the micro-calorimeter, will enable a wide variety of important science themes to be pursued. 
ASTRO-H is expected to provide breakthrough results in scientific areas as diverse as the large-scale structure of the Universe and its evolution, the behavior of matter in the gravitational strong field regime, the physical conditions in sites of cosmic-ray acceleration, and the distribution of dark matter in galaxy clusters at different redshifts.

\end{abstract}




\section{Introduction}

The history and evolution of the Universe can be described as a process in which structures 
of different scales,  such as stars, galaxies, and clusters of galaxies, are formed.
In some cases, during this process, the matter and energy concentrate to an extreme degree in the form 
of black holes and neutron stars.  It is a mystery of Nature why and how  the
overwhelming diversity over orders of magnitude  in spatial and density scales has been produced in the Universe
following an expansion from a nearly uniform state. Excellent  probes of this process are clusters of galaxies, the largest astronomical objects in
the Universe. Observing clusters of galaxies and revealing their history will lead to an understanding
of how the largest structures form and evolve in the Universe. Equally important is  studying how supermassive black holes form and develop, and what a role they play in the evolution of galaxies and clusters of galaxies.

The X-ray band is capable of probing extreme environments of the Universe such as those near black holes or the surface of neutron stars,
as well as observing  the emission from high temperature gas and tracing accelerated electrons.
In recent years, \chandra, \xmm, \suzaku\ and other X-ray missions have made great advances in X-ray Astronomy. We have obtained knowledge that has revolutionized our understanding of the high energy Universe and we have learned that phenomena observed in the X-ray band are deeply connected to those observed in other wavelengths from radio to $\gamma$-rays.

 In order to revolutionize X-ray astronomy even further, the ASTRO-H mission has been designed and is currently being constructed.
ASTRO-H is an international X-ray satellite 
that Japan plans to launch with the H-II A rocket in 2015\cite{Ref:Proposal,Ref:Proposal03,Ref:Kunieda2004,Ref:Takahashi,Ref:Takahashi2008,Ref:Takahashi2010,Ref:Takahashi2012}.  
NASA has selected the US participation in  ASTRO-H as a Mission of Opportunity in the Explorer Program category.   
Under this program, the NASA/Goddard Space Flight Center collaborates with 
ISAS/JAXA on the implementation of an X-ray micro-calorimeter and soft X-ray telescopes
(SXS Proposal NASA/GSFC, 2007)\cite{Ref:ProposalNASA}. Other institutional members of the collaboration,
building hardware for ASTRO-H, are SRON, Geneva University, CEA/DSM/IRFU, CSA,  Stanford 
University, and ESA.
In early 
2009, NASA, ESA and JAXA have selected science advisors to provide 
scientific guidance to the ASTRO-H project regarding the design/development 
and operation phases of the mission. The ESA contribution to the ASTRO-H Mission includes the procurement of payload hardware elements 
that enhance the scientific capability of the mission.

In this paper, we describe the ASTRO-H satellite and report the recent progress of the construction of the satellite and the mission instruments,  by updating our previous paper published in 2012\cite{Ref:Takahashi2012}.

\section{Science Requirements}

The prime scientific goals for ASTRO-H are fundamental questions in contemporary astrophysics, as listed below.

\begin{center}
\bf{Scientific Goals and Objectives}
\end{center}

\subsubsection*{Revealing the large-scale structure of the Universe and its evolution}

\begin{itemize}
\item 
ASTRO-H will observe clusters of galaxies, the largest bound structures in the Universe, with the aim
to reveal the interplay between the thermal energy of the intra-cluster medium and the kinetic energy
of sub-clusters, from which clusters form; measure their non-thermal energy and chemical composition; and to directly trace the
dynamic evolution of clusters of galaxies.

\item ASTRO-H will observe distant supermassive black holes hidden 
by thick intervening material  with  100 times higher sensitivity 
than the currently operating Suzaku satellite, and will study their evolution and the role they play in galaxy formation.
\end{itemize}

\subsubsection*{Understanding the extreme conditions in the Universe}

\begin{itemize}
\item ASTRO-H will measure the motion of matter very close to black 
holes with the aim to sense the gravitational distortion of  space,  to understand 
the structure of relativistic space-time and to study the physics of accretion processes.
\end{itemize}

\subsubsection*{Exploring the diverse phenomena of the non-thermal Universe}
\begin{itemize}
\item ASTRO-H will derive the physical conditions of the sites where 
high energy cosmic ray particles gain energy and will elucidate 
the processes by which gravity, collisions, and stellar explosions energize  
those cosmic rays.

\end{itemize}

\subsubsection*{Elucidating dark matter and dark energy}
\begin{itemize}
\item ASTRO-H will map the distribution of dark matter in clusters 
of galaxies and will determine the total mass of galaxy clusters 
at different distances (and thus at different ages), and will study the role 
of dark matter and dark energy in the evolution of these systems.

\end{itemize}

In order to achieve the cutting-edge scientific goals
described above, ASTRO-H is designed with  the following features:
\begin{enumerate}
\item High resolution spectroscopy of extended objects  with X-ray micro calorimeters;
\item Hard X-ray imaging up to 80 keV using multi-layer coatings on X-ray telescopes;
\item Soft X-ray Imaging with large field of view of 38$\times$38 arcmin$^2$;
\item Wide coverage in energy up to 600 keV using a narrow-field Compton camera.
\end{enumerate}

\section{Spacecraft}

\begin{figure}
\hspace{2cm}\includegraphics[scale=0.35]{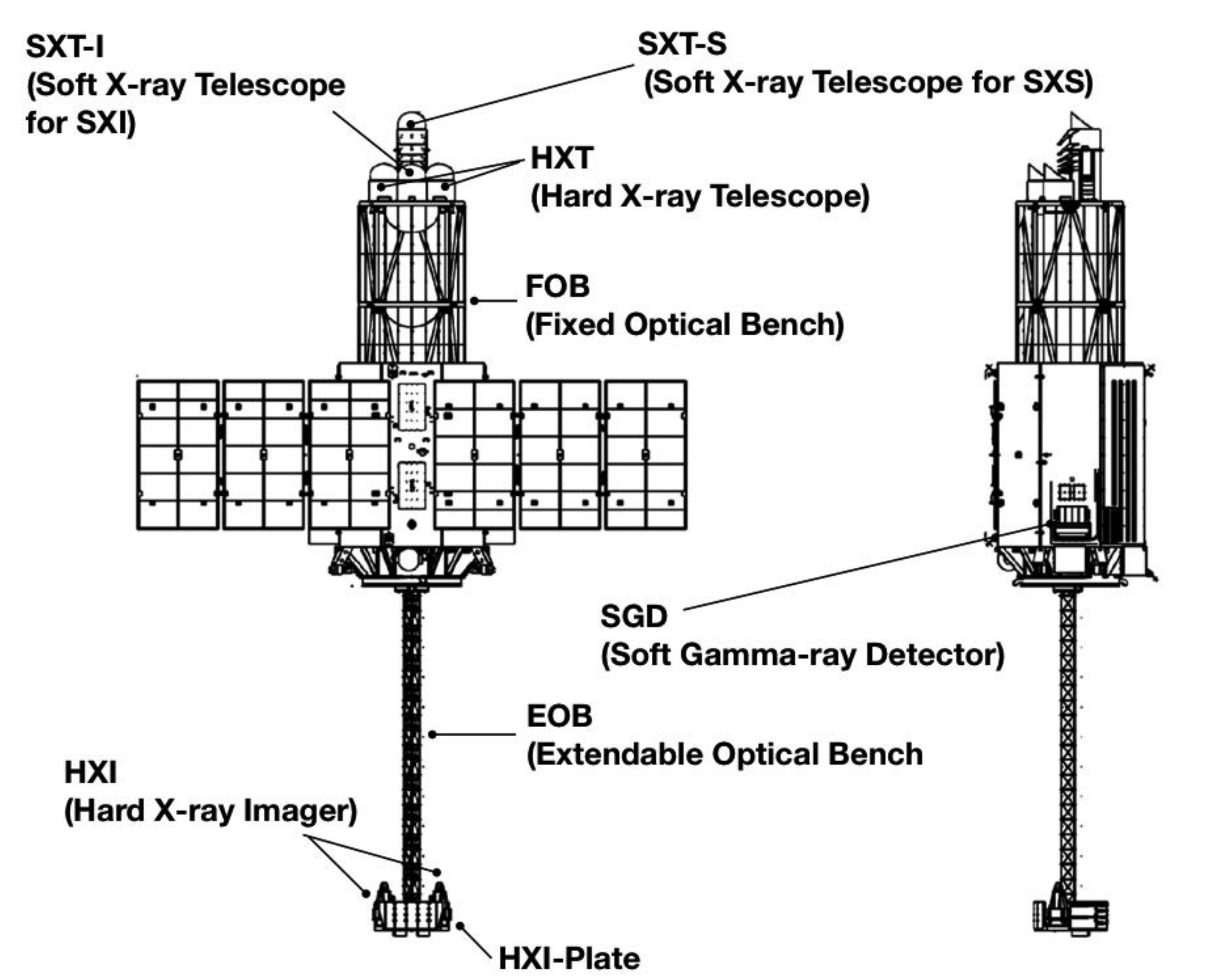}
\caption{Schematic view of the \astroh\ satellite with the Extendable Optical Bench deployed. }
\label{Fig:ASTRO-H-1}
\end{figure}

\begin{figure}
\centerline{\includegraphics[width=10cm]{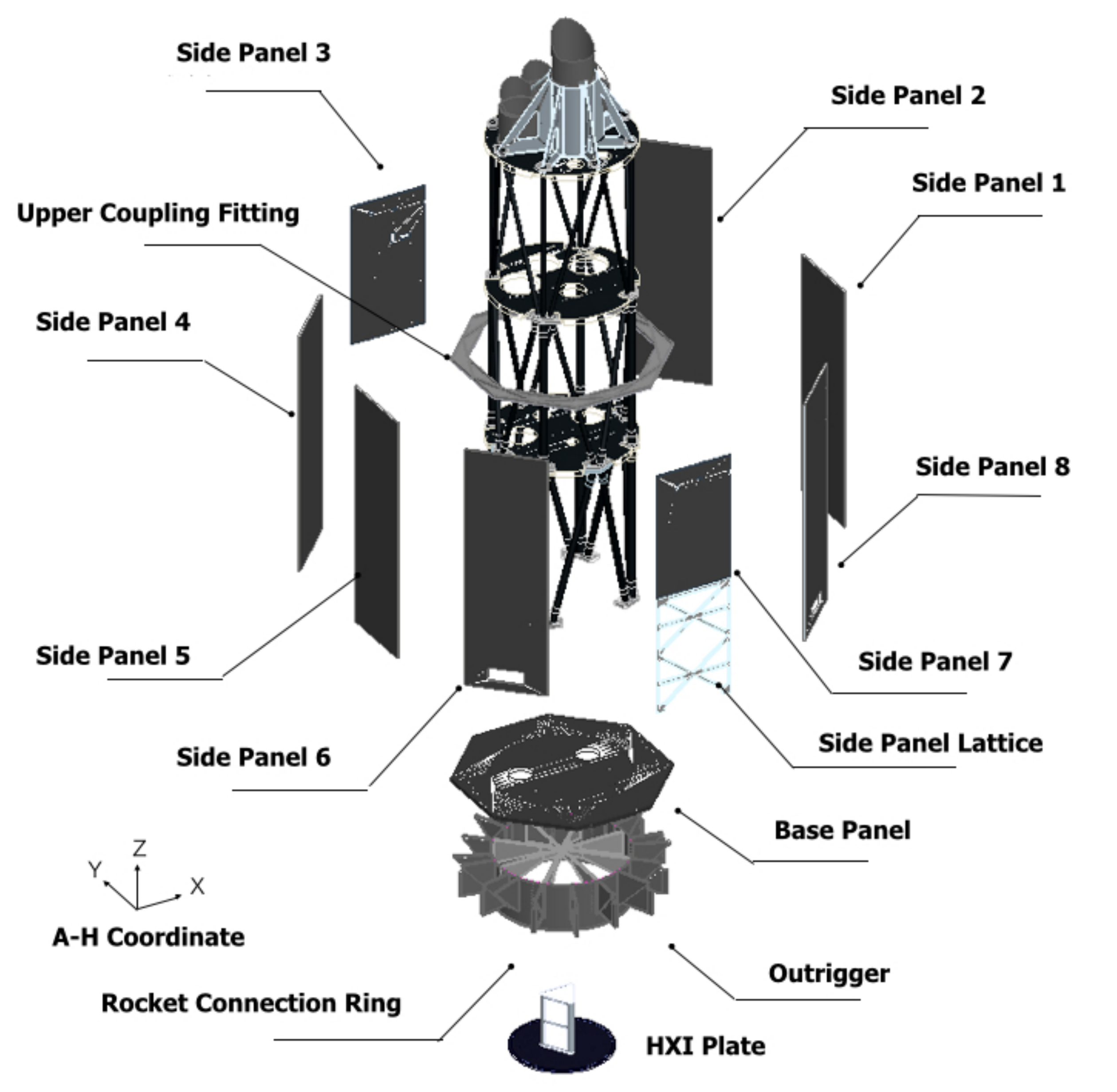}}
\caption{Exploded view of the structure  of ASTRO-H. }
\label{Fig:ASTRO-H-2}
\end{figure}

\begin{table}
\caption{ASTRO-H Mission}
\begin{center}
\begin{tabular}{ll}
\hline
 Launch site & 
Tanegashima Space Center, Japan\\
 Launch vehicle& JAXA H-IIA rocket\\
 Orbit Altitude& $\sim$550 km\\
 Orbit Type& Approximate circular orbit\\
 Orbit Inclination& $\sim$ 31 degrees\\
 Orbit Period& 96 minutes\\
Total Length& 14 m\\
 Mass& $\sim$ 2.7  metric ton\\
 Power& $<$ 3500 W\\
 Telemetry Rate& 8 Mbps (X-band QPSK) \\
 Recording Capacity& 12 Gbits at EOL\\
 Mission life & $>$ 3 years\\
\hline
 \end{tabular}
 \end{center}
\end{table}

There are four focusing telescopes mounted on the top of a fixed optical bench (FOB). Two of the four telescopes are Soft X-ray Telescopes (SXTs) and they have a 5.6 m focal length.  They will focus medium-energy X-rays (E $\sim$ 0.3--12 keV) onto focal plane detectors mounted on the base plate of the spacecraft (see Figs \ref{Fig:ASTRO-H-1} and \ref{Fig:ASTRO-H-2}). 
  One SXT will point to a micro-calorimeter spectrometer array with
excellent energy resolution of $\leqq$7 eV, and the other SXT will point to
a large-area CCD array. 
The other two telescopes are Hard X-ray Telescopes (HXTs) capable of focusing high-energy X-rays (E = 5--80 keV).  The focal length of the HXTs is 12 m.   The Hard X-ray Imaging detectors (HXIs) are mounted on the HXI plate, at the end of a 6 m extendable optical bench (EOB) that is stowed to fit in the launch fairing and deployed once in orbit.  
 In order to extend the energy 
coverage to the soft $\gamma$-ray region up to 600 keV, the Soft Gamma-ray 
Detector (SGD) will be implemented as a non-focusing detector.  
  Two SGD detectors, each consisting of three units
 will be mounted separately on two sides of the satellite. With these instruments, 
ASTRO-H will cover the entire bandpass between 0.3~keV and 600~keV.  The key parameters  of 
those instruments  are 
summarized in Table.~\ref{Table:Spec}.

The lightweight design of the EOB renders it vulnerable to distortions from thermal fluctuations in low-Earth orbit (LEO) and spacecraft attitude maneuvers.  Over the long exposures associated with X-ray observing, such fluctuations might impair HXI image quality unless a compensation technique is employed. To provide the required corrections, the Canadian contribution to the the ASTRO-H project is  a laser metrology system (the Canadian ASTRO-H Metrology System, CAMS) that will measure
displacement in the alignment of the HXT optical path. The CAMS consist of 
 a laser and detector module (CAMS-LD) located on the top plate of the FOB, and a passive target module (CAMS-T) consisting of a retroreflector (corner cube mirror) mounted on the EOB detector plate (HXI plate)\cite{Ref:Gallo,Ref:Gallo2014}.

The spacecraft attitude is stabilized by four sets of reaction wheels with one redundancy, while the attitude is measured by two star trackers and its change rate by two gyroscopes. There are two more gyroscopes mounted in skew directions, which provide redundancy. The accumulated angular momentum is unloaded by magnetic torquers that interact with the Earth's magnetic field.  The required accuracy of the spacecraft attitude solution is approximately 0.$^\prime$33  with a stability of better than 0.$^\prime$12 per 4s (a nominal exposure time for the CCDs).
The pointing direction of the telescope is limited by the power constraint of the solar panel. The area of the sky accessible at any time is
a belt within which the Sun angle is between 60$^\circ$ and 120$^\circ$. Any part of the sky is accessible at least twice a year.  It is expected to take $\sim$72 min for  a 180$^\circ$ maneuver.

Almost all of onboard subsystems of ASTRO-H, such as the command/data handling system, the attitude control system, and four types of X-ray/gamma-ray telescope instruments, are connected to the SpaceWire network using a highly redundant topology\cite{Ref:Yuasa}.
 The number of physical SpaceWire links between components exceeds 140 connecting $\sim$40 separated components (i.e., separated boxes), and there are more links in intra-component (intra-board) networks. Most of the electronics boxes of both the spacecraft bus and the scientific instruments are mounted on the side panels of the space craft.
The electronics boxes for the HXI are mounted on the HXI plate. 

%
%

\section{Science Instruments}

\begin{figure}[htb]
\centerline{\includegraphics[height=7cm]{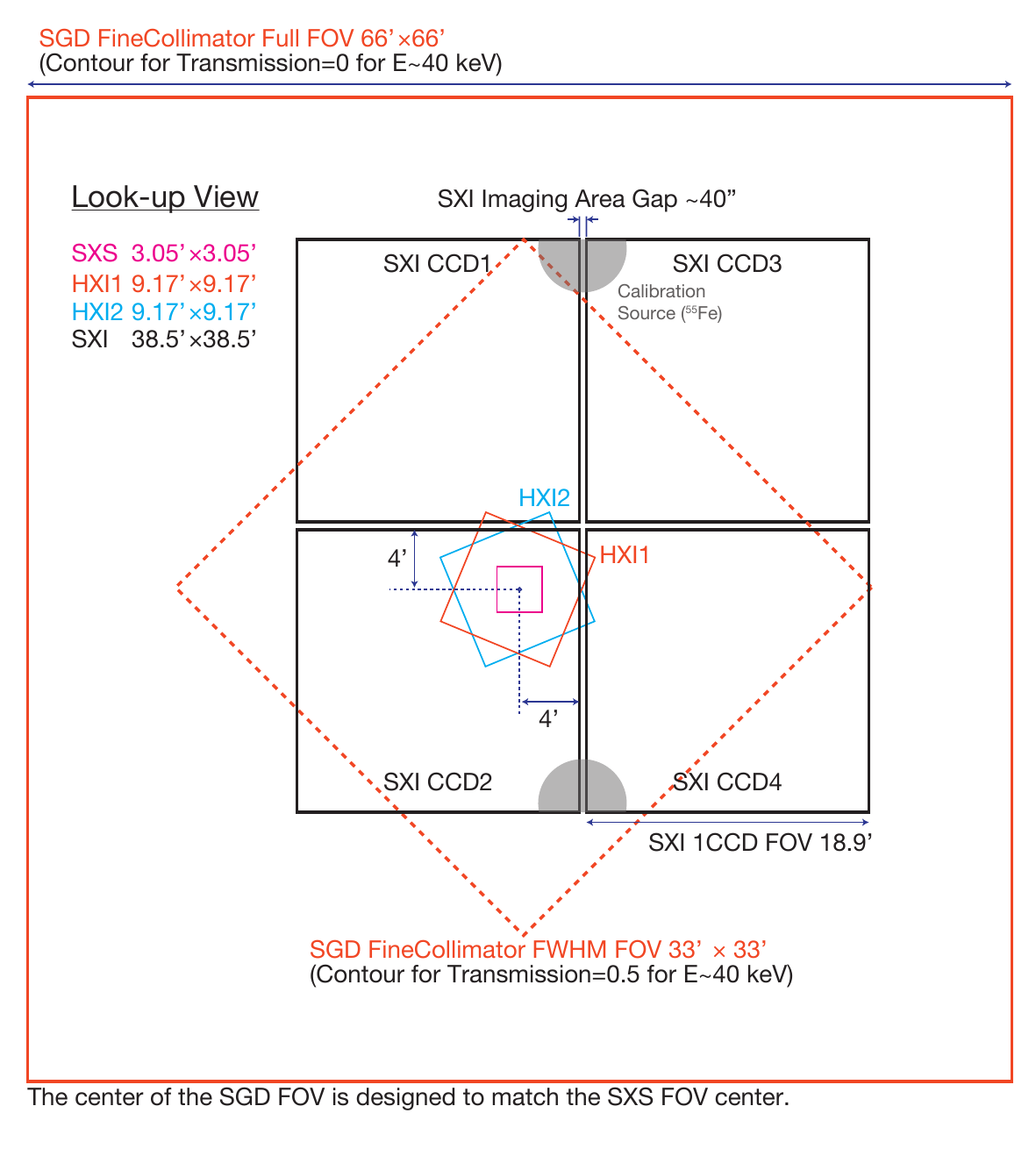}\includegraphics[height=7cm,clip]{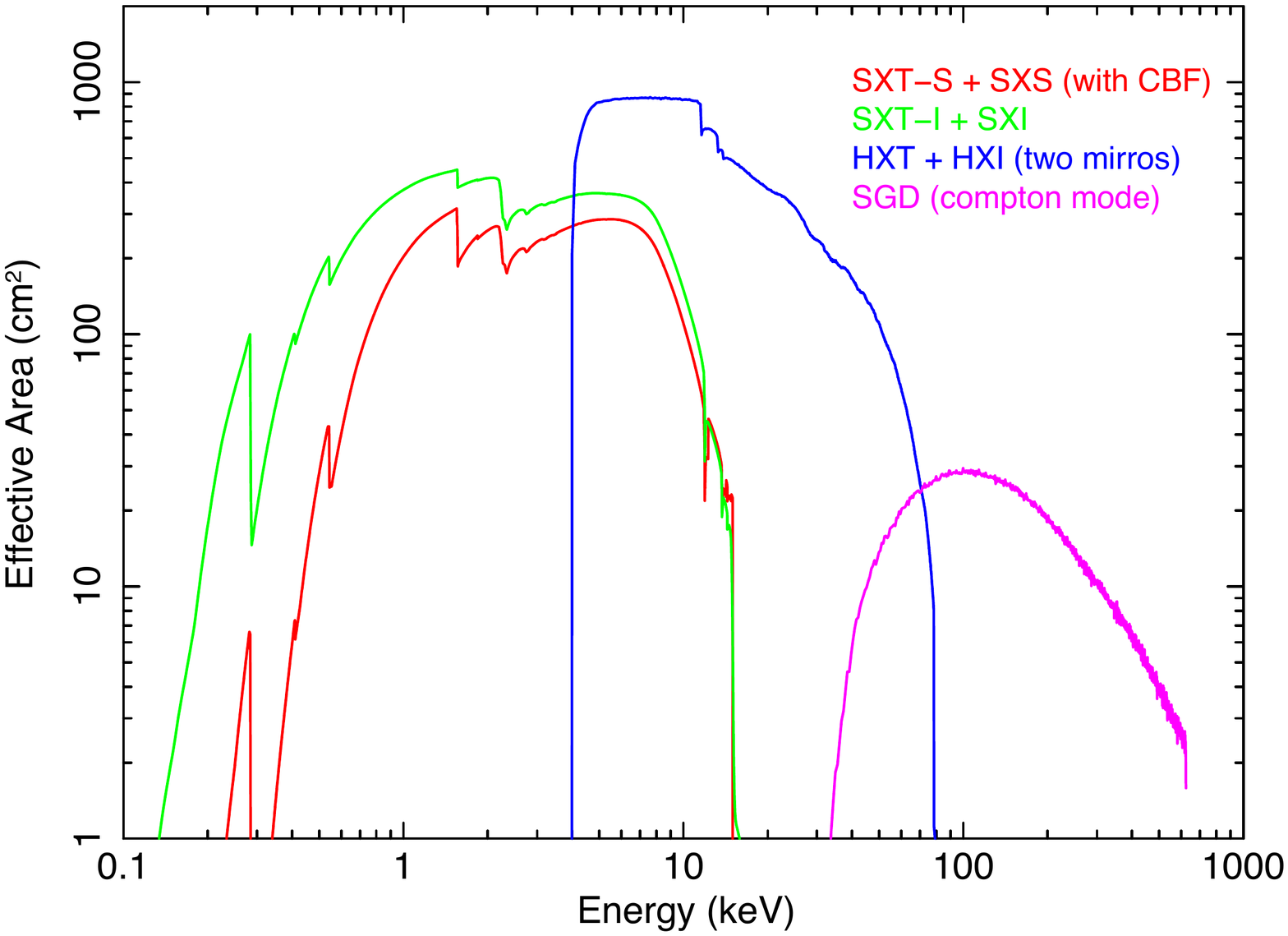}}
\caption{(left) Fields of view of the ASTRO-H instruments, SXS, SXI and HXI. FWHM FOV of a SGD fine collimator is also shown. (right) Effective areas of the ASTRO-H instruments, SXS, SXI and HXI, combined with the telescopes. Effective area of SGD is also shown, when the Compton mode is used. }
\label{Fig:FOVArea}
\end{figure}

ASTRO-H instruments include a high-resolution, high-throughput spectrometer 
sensitive over 0.3--12~keV with high spectral resolution of 
$\Delta E $ $\leqq$ 7~eV, enabled by a micro-calorimeter 
array located in the focal plane of thin-foil X-ray optics;  
hard X-ray imaging spectrometers covering 5--80~keV, located in 
the focal plane of multilayer-coated, focusing hard X-ray mirrors;  
a wide-field imaging spectrometer sensitive over 0.4--12~keV, 
with an X-ray CCD camera in the focal plane of a soft X-ray 
telescope;  and a non-focusing  Compton-camera 
type soft gamma-ray detector, sensitive in the 40--600~keV band. 
The FOVs and effective areas of these instruments are shown in Fig. \ref{Fig:FOVArea}. The simultaneous 
broad bandpass, coupled with high spectral resolution, will enable 
the pursuit of a wide variety of important science themes.  

In the following sections, these instruments  are briefly described.
Detailed descriptions of the instruments and their current status are available  in other papers in these\cite{Ref:Soong2014,Ref:Awaki2014,Ref:Mitsuda2014,Ref:Sato2014,Ref:Hayashida2014,Ref:Fukazawa2014}. 
 and previous proceedings of this conference.
\cite{Ref:Okajima2012,Ref:Awaki2012,Ref:Fujimoto2012,Ref:Tsunemi2012,Ref:Kokubun2012,Ref:Watanabe2012,Ref:FW2012}.

\begin{table}
\caption{Key parameters of the ASTRO-H payload}
\begin{center}
\label{Table:Spec}
\begin{footnotesize}
\begin{tabular}{|l|p{2.5cm}|p{2.5cm}|p{2.4cm}|p{2.8cm}|}
\hline
Parameter & Hard X-ray  & Soft X-ray & Soft X-ray   & Soft $\gamma$-ray  \\
&  Imager & Spectrometer  &  Imager & Detector\\
&   (HXI) & (SXS)  &   (SXI) & (SGD)\\
\hline
\hline
Detector & Si/CdTe & micro& X-ray & Si/CdTe  \\
 technology & cross-strips & calorimeter & CCD & Compton Camera \\
 \hline
Focal length & 12 m & 5.6 m & 5.6 m  & -- \\
\hline
Effective area & 300 cm$^{2}$@30 keV  &  210 cm$^{2}$@6 keV  &  360 cm$^{2}$@6 keV & 
 $>$20~cm$^{2}$@100 keV  \\
 &  & 160 cm$^{2}$ @ 1 keV  &    &  Compton Mode  \\
\hline
Energy range & 5 --80 keV & 0.3 -- 12 keV  & 0.4 -- 12 keV & 40 -- 600 keV \\
\hline
Energy  & 2 keV& $<$ 7 eV & $<$ 200 eV& $<$ 4 keV \\
resolution& (@60 keV)  & (@6 keV) & (@6 keV) &  (@60 keV) \\
 (FWHM) & &  & & \\
 \hline
Angular  & $<$1.7 arcmin& $<$1.3 arcmin& $<$1.3 arcmin& -- \\
resolution & & & & \\
\hline
Effective & $\sim$ 9 $\times$ 9  & $\sim$ 3 $\times$ 3   & $\sim$ 38 $\times$ 38 & 0.6 $\times$ 0.6 deg$^{2}$ \\
 Field of View &arcmin$^{2}$ & arcmin$^{2}$& arcmin$^{2}$& ($<$ 150 keV) \\
\hline
Time resolution  & 25.6  $\mu$s & 5  $\mu$s &  4 sec/0.1 sec & 25.6  $\mu$s \\
\hline
Operating  & $-$20$^{\circ}$C & 50 mK & $-$120$^{\circ}$C& $-$20$^{\circ}$C\\
 temperature  & & & & \\
 \hline
\end{tabular}
\end{footnotesize}
\end{center}
\end{table} 

\subsection{Soft X-ray Telescopes}

The X-ray mirror is very similar to the Suzaku X-ray Telescope, but with a longer focal length of 5.6 m and a larger outer diameter of 45 cm. 
The SXT consists of three parts, an X-ray mirror, a stray light baffle called the pre-Collimator,
and a thermal shield to keep the mirror temperature at around 20\degree.
The mirror is  conically approximated Wolter I grazing incidence
optic with 203 nested shells. Each shell is segmented into four quadrants\cite{Ref:Soong2014,Ref:Okajima2012,Ref:Okajima,Ref:Selemistos2010}

\begin{figure}
\centerline{\includegraphics[width=14cm]{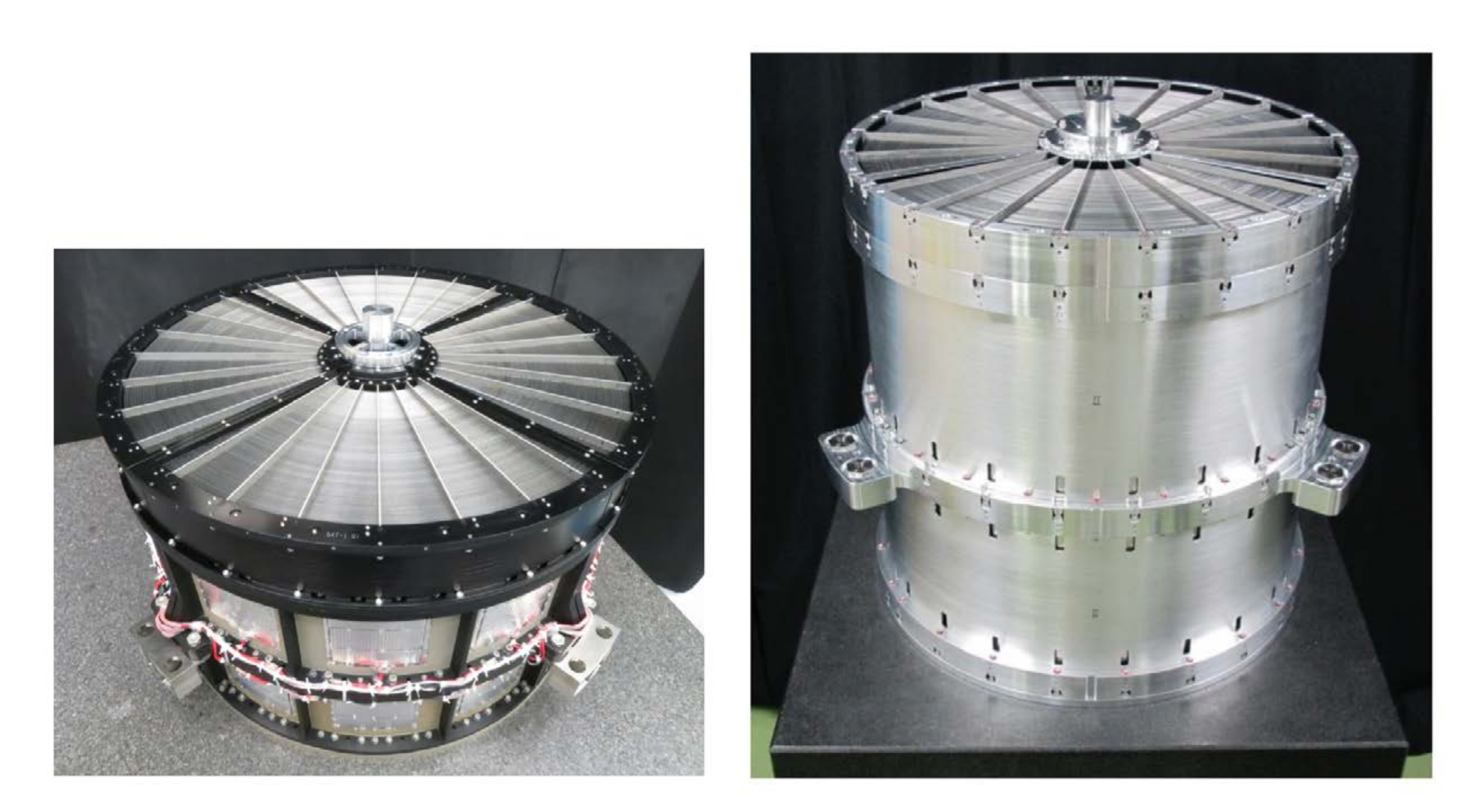}}
\caption{Photographs of flight models of (left) Soft X-ray telescope, SXT-2 and (right) hard X-ray telescope, HXT-1.}
\label{Fig:XRTFM}
\end{figure}

The flight SXTs (Fig. \ref{Fig:XRTFM} left) were fabricated at NASA/GSFC and have been delivered to JAXA.
According to calibration at GSFC and ISAS, the angular resolution 
 (Half Power Diameter : HPD) is 1.3 arcmin and 1.2 arcmin for
 SXT-1 and SXT-2, respectively.  The result obtained with  SXT-2 exceeds the desired goal.
 Effective areas are measured to be
 $\sim$590 cm$^2$ at 1keV and $\sim$430 cm$^2$ at 6 keV.In order to collect more photons, 
SXT-2, which has  slightly better angular resolution, will be used for the SXS, because  it  has a smaller detector area.
Ground calibration for SXT-1 and SXT-2 has been completed. All the basic performance characteristics, effective area (on/off-axis), PSF (on/off-axis), and stray light have been measured during the calibration.

\subsection{Hard X-ray Telescopes}

Most non-imaging X-ray instruments flown so far were essentially limited to
studies of sources with 10--100~keV fluxes of  $>$4 $\times$ 10$^{-12}$--%
10$^{-11}$~erg~cm$^{-2}$s$^{-1}$, at best. The exception is NASA's NuSTAR mission\cite{Ref:NuSTARApJ}, launched in June 2012, which has a multi-layer-coated focusing hard X-ray telescope similar to the HXT
This limitation is due to the presence
of high un-rejected backgrounds from particle events and cosmic X-ray
radiation, which increasingly dominate above 10~keV.
Imaging \em{-} and especially focusing \em{-} instruments Imaging, and
especially focusing instruments have two tremendous advantages. Firstly,
the volume of the focal plane detector can be made much smaller than for
non-focusing instruments, thus reducing the absolute background level
since the background flux generally scales with the size of the
detector.  Secondly, the residual background, often time-variable, can
be measured simultaneously with the source, and can be reliably
subtracted.  For these reasons, a focusing hard X-ray telescope in
conjunction with an imaging detector sensitive for hard
X-ray photons is the appropriate choice to achieve a breakthrough in
sensitivity for the field of high energy astronomy.

A depth-graded  multi-layer mirror reflects X-rays not only by total
external reflection but also by Bragg reflection. In order to obtain
a high reflectivity up to 80~keV, the HXTs consist of a stack of
multilayers with different sets of periodic length and number of layer
pairs with a platinum/carbon coating. The technology of a hard X-ray
focusing mirror has already been proven by the balloon programs
InFOC$\mu$S (2001, 2004)\cite{Ref:Kunieda2006,Ref:Ogasaka:In}, 
HEFT (2004)\cite{Ref:Fiona} and SUMIT (2006)\cite{Ref:Kunieda2006}
and very recently with the NuSTAR satellite\cite{Ref:NuSTARApJ}.

The HXT  consists of three parts, an X-ray mirror:  a stray light baffle called pre-Collimator,
 and a thermal shield (Fig. \ref{Fig:XRTFM}, right).
The  mirror is  based on conically-approximated Wolter I grazing incidence
optics\cite{Ref:Awaki2014,Ref:Awaki2012}. The diameters of the innermost and the outermost reflectors are 120 mm
and 450 mm, respectively. The total number of the nested shells is 213 per  quadrant and 
 thus 1278 reflectors in each telescope. 
Production of  two flight-ready HXTs, HXT-1 and HXT-2, has been completed in 2014.
According to  calibration performed by using the SPring-8 beam line,  
the characteristics of  HXT1 and HXT2 are quite similar.
Their angular resolution  (HPD) is $\sim$1.9 arcmin at 30 keV.
With a focal length of 12 m,   a collecting area of 174 cm$^2$ at 30 keV  for one telescope
has been achieved, resulting in a total effective area of 348 cm$^2$.

\subsection{Soft X-ray Spectrometer System}

The soft X-ray Spectrometer (SXS) consists of the Soft X-ray Telescope 
(SXT), the filter wheel (FW) assembly\cite{Ref:FW2012}, the X-ray Calorimeter Spectrometer (XCS) and the cooling system\cite{Ref:Mitsuda,Ref:Mitsuda2010,Ref:Mitsuda2014}.
The XCS is a 36-pixel system with an energy resolution of $\leq$7~eV between 0.3--12~keV. 
The array design for the SXS is basically the same as that for the Suzaku/XRS, but has larger pixel pitch and absorber size.  HgTe absorbers are attached to ion-implanted Si thermistors formed on suspended Si micro-beams.\cite{Ref:Kelley,Ref:Porter1,Ref:Porter2}  The array has 814 $\mu$m pixels on an 832 $\mu$m pitch and was manufactured during the Suzaku/XRS program along with arrays with smaller pixel size as an option for a larger field of view.  
For ASTRO-H, the longer focal length of the SXS (5.6 m vs. 4.5 m for Suzaku) necessitated the use of these larger arrays to maintain a FOV of at least 2.9 $\times$ 2.9 arcminutes.  The 8.5-micron-thick absorbers were fabricated by EPIR Corporation and diced using reactive ion etching.  These  absorbers provide high quantum efficiency across the 
0.3--12~keV band. Despite the larger pixel size for the SXS (factor of 1.7 in volume), the energy resolution is substantially improved from $\sim$ 6 eV to $\sim$ 4 eV (FWHM).  The main reasons for this are that EPIR developed a process to produce HgTe with lower specific heat and that the operating temperature of the instrument has been lowered from 60 mK to 50 mK\cite{Ref:Porter2}.

\begin{figure}
\includegraphics[width=16cm]{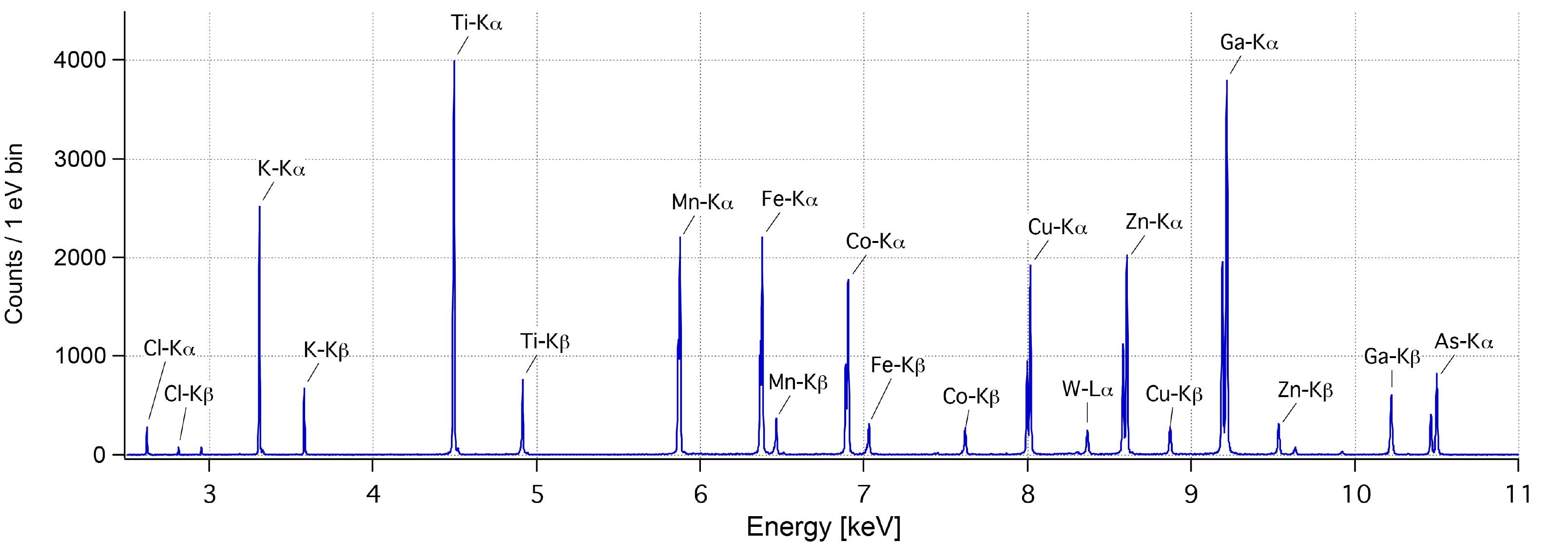} 
\caption{Energy spectrum of the flight array obtained with a series of fluorescent targets used to determine the energy scale.  The energy resolution of all 36 pixels combined is $\sim$4.7 eV FWHM as measured at 5.9 keV.}
\label{Fig:FMSpec}
\end{figure}

The SXS uniquely performs high-resolution spectroscopy of extended sources. 
In contrast to a grating, the spectral resolution of the calorimeter is 
unaffected by the source angular size because it is non-dispersive.  
Figure \ref{Fig:FMSpec} shows an energy spectrum taken with the flight sensor array.
For all sources with angular extent larger than 30~arcsec, Chandra MEG 
energy resolution is degraded compared with that of a CCD; the energy 
resolution of the XMM-Newton RGS is similarly degraded for sources 
with angular extent $\geq$25 arcsec. SXS therefore makes possible high-resolution 
spectroscopy of sources inaccessible to current grating instruments.

\begin{figure}
\centerline{\includegraphics[width=10cm]{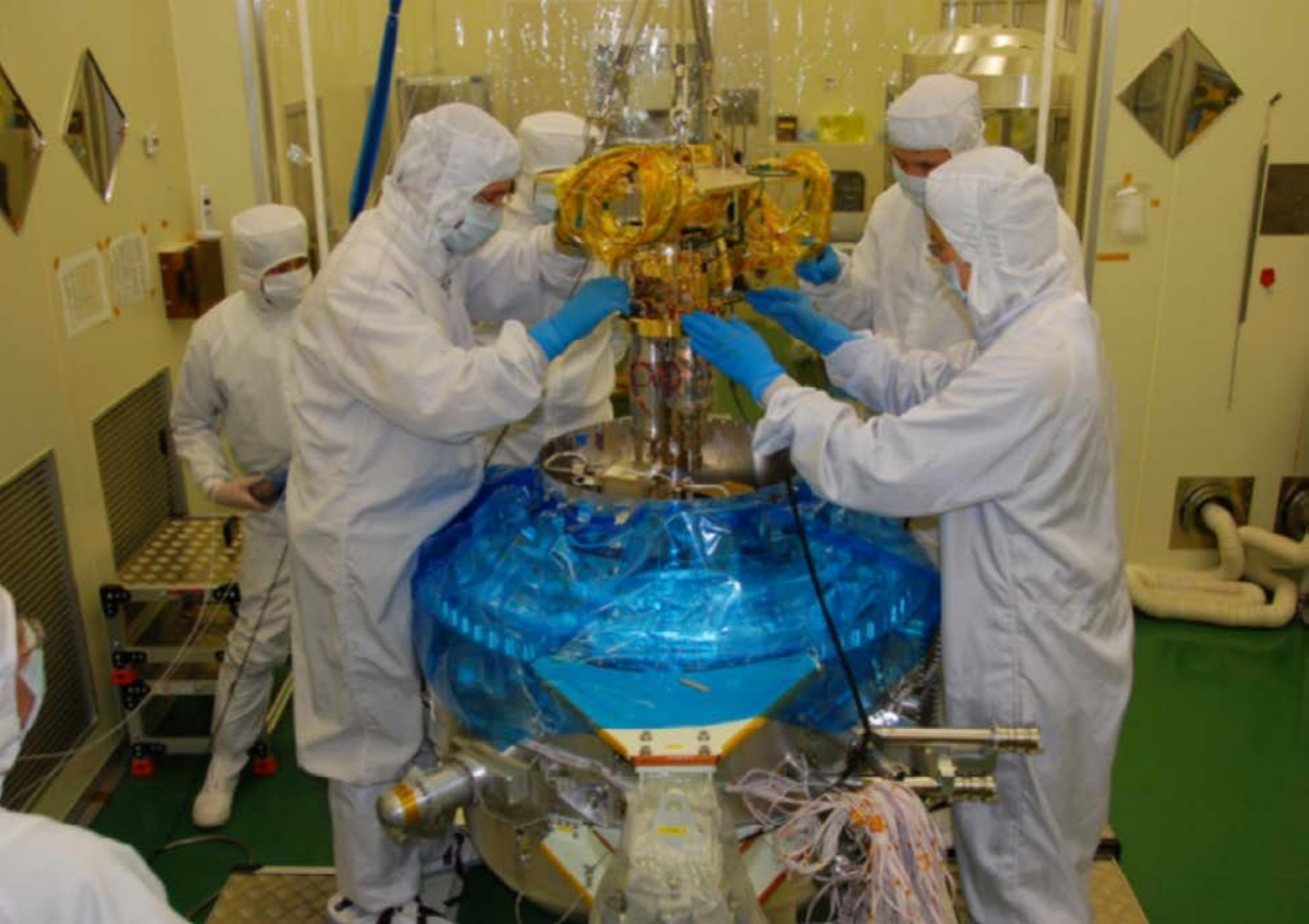} }
\caption{Installation of the flight calorimeter sensor module into the Dewar.}
\label{Fig:SXSInstall}
\end{figure}

With a 5.6-m focal length, the 0.83~mm pixel pitch corresponds to 
0.51~arcmin, giving the array a field of view of 3.05~arcmin on a side.  
The detector assembly provides electrical, thermal, and mechanical interfaces 
between the detectors (calorimeter array and anti-coincidence particle detector) 
and the rest of the instrument.
The SXS effective 
area at 6~keV will be at least 210~cm$^{2}$, a 60~\% increase over the Suzaku XRS,  while at 1~keV 
the SXS has 160~cm$^{2}$, a 20~\% increase.

The XCS cooling system must cool the array to 50~mK with sufficient duty cycle to 
fulfill the SXS scientific objectives:  this requires extremely low heat loads.  
To achieve the necessary gain stability and energy resolution, the cooling 
system must regulate the detector temperature to within 2~$\mu$K rms for at 
least 24~hours per cycle (see Fig. \ref{Fig:SXSInstall})\cite{Ref:Fujimoto2010}.  From the detector stage to room temperature, the 
cooling chain is composed of a 3-stage  Adiabatic Demagnetization 
Refrigerator (ADR), superfluid liquid $^{4}$He (hereafter LHe), 
a $^{4}$He Joule-Thomson (JT) cryocooler, and two-stage Stirling cryocoolers.  
An ADR has been adopted because it readily meets 
the requirements for detector temperature, stability, recycle time, 
reliability in the space environment, and previous flight heritage\cite{Ref:Shirron}.  The 
design of Stirling cryocoolers is based on coolers developed for space-flight 
missions in Japan (Suzaku, AKARI, and the SMILES instrument  deployed on 
the ISS\cite{Ref:Narasaki}) that have achieved an excellent performance with 
respect to cooling power, efficiency, long life and mass.  Thirty litres of LHe are  used
as a heat-sink for the 2-stage 
ADR. To reduce the parasitic heat load on the He tank, a 
$^{4}$He JT cryocooler is used to cool a 4 K shield. To achieve redundancy for failure 
(unexpected loss) of LHe, another  ADR (3rd stage ADR) is used between the 
He tank and the JT cryocooler, with two heat-switches on both sides.  
This ADR is operated if LHe is lost, to cool down the 1 K shield (He tank). 
\red{A series of five blocking filters shield the calorimeter array from UV and longer wavelength radiation.  Two of these are contained within the detector assembly and are free-standing aluminized polyimide, essentially the same as successfully used on Suzaku.  The remaining three are installed on two of the three vapor-cooled shields, and one on the dewar main shell.  These filters also have aluminized polyimide, but are supported by two-level, high-throughput, Si meshes.  The meshes provide increased strength, and enable heaters on the perimeter of the filters that can be used to drive off contamination if necessary.}

\begin{figure}
\includegraphics[width=16cm]{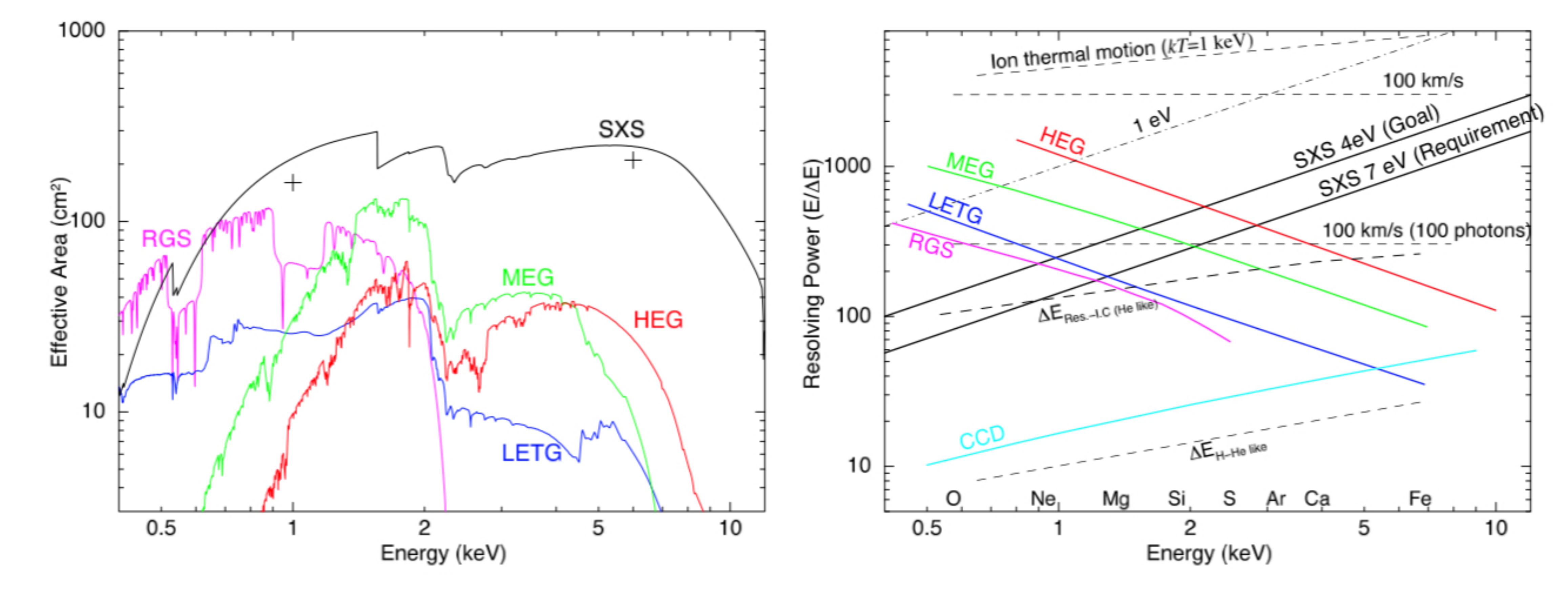}  
\caption{(left) Effective areas of high-resolution X-ray spectroscopy missions as functions of X-ray energy. The curve for the \astroh\ SXS is the present best estimate for a point source. The two crosses show the mission requirements at specific energies. The \xmm\ RGS effective area is the summation of first order spectra of the two instruments (RGS-1 and RGS-2). The effective areas of the LETG, MEG and HEG onboard \chandra\ are summations of the first order dispersions in $\pm$ directions.
(right) Resolving power of the \astroh\ SXS as a function of X-ray energy for the two cases, 4\,eV resolution (goal) and 7\,eV (requirement). The resolving power of high resolution instruments onboard \chandra\ and \xmm\ and typical resolving power of X-ray CCD cameras are also shown for comparison \cite{Ref:Mitsuda2010}. }
\label{Fig:XRS}
\end{figure}

In combination with a high throughput X-ray telescope, the 
SXS improves on the Chandra and XMM-Newton grating spectrometers in 
two important ways. At E $>$ 2~keV, SXS is both more sensitive and has 
higher resolution (Fig. \ref{Fig:XRS}), especially in the Fe K band where 
SXS has 10 times larger  collecting area and much better energy resolution, giving a net 
improvement in sensitivity by a factor of 30 over Chandra.  The broad bandpass of 
the SXS encompasses the critical emission and absorption lines 
of Fe I-XXVI between 6.4 and 9.1~keV. Fe K lines provide particularly useful 
diagnostics because of their (1) strength, due to the high abundance and large 
fluorescent yield (30\%), (2) spectral isolation from other lines, and (3) 
relative simplicity of the atomic physics. Fe K emission lines reveal conditions 
in plasmas with temperatures between 10$^{7}$ and 10$^{8}$ K, which are typical values for 
stellar accretion disks, SNRs, clusters of galaxies, and many stellar coronae.  
In cooler plasmas, Si, S, and Fe fluorescence and recombination occurs 
when an X-ray source illuminates nearby neutral material. Fe emission 
lines provide powerful diagnostics of non-equilibrium ionization 
due to inner shell K-shell transitions from Fe XVII--XXIV\cite{Ref:Decaux}.

In order to obtain a good performance for bright sources, 
a filter wheel (FW) assembly, which includes a wheel with selectable filters and a set of modulated X-ray sources,
 are provided by SRON and Univ. of Geneva.
This is placed at a distance of 90 cm from the detector. The FW is able to rotate a suitable filter into the
beam to optimize the quality of the data, depending on the source characteristics\cite{Ref:Vries,Ref:FW2012}. In addition to the
filters, a set of on-off-switchable X-ray calibration sources, using a light sensitive photo-cathode, will be implemented.
With these calibration sources, it is possible to calibrate the energy scale with a typical 1$-$2 eV accuracy, and 
 will allow proper gain and linearity calibration of the detector in flight.

\begin{figure}
\centerline{\includegraphics[width=12cm]{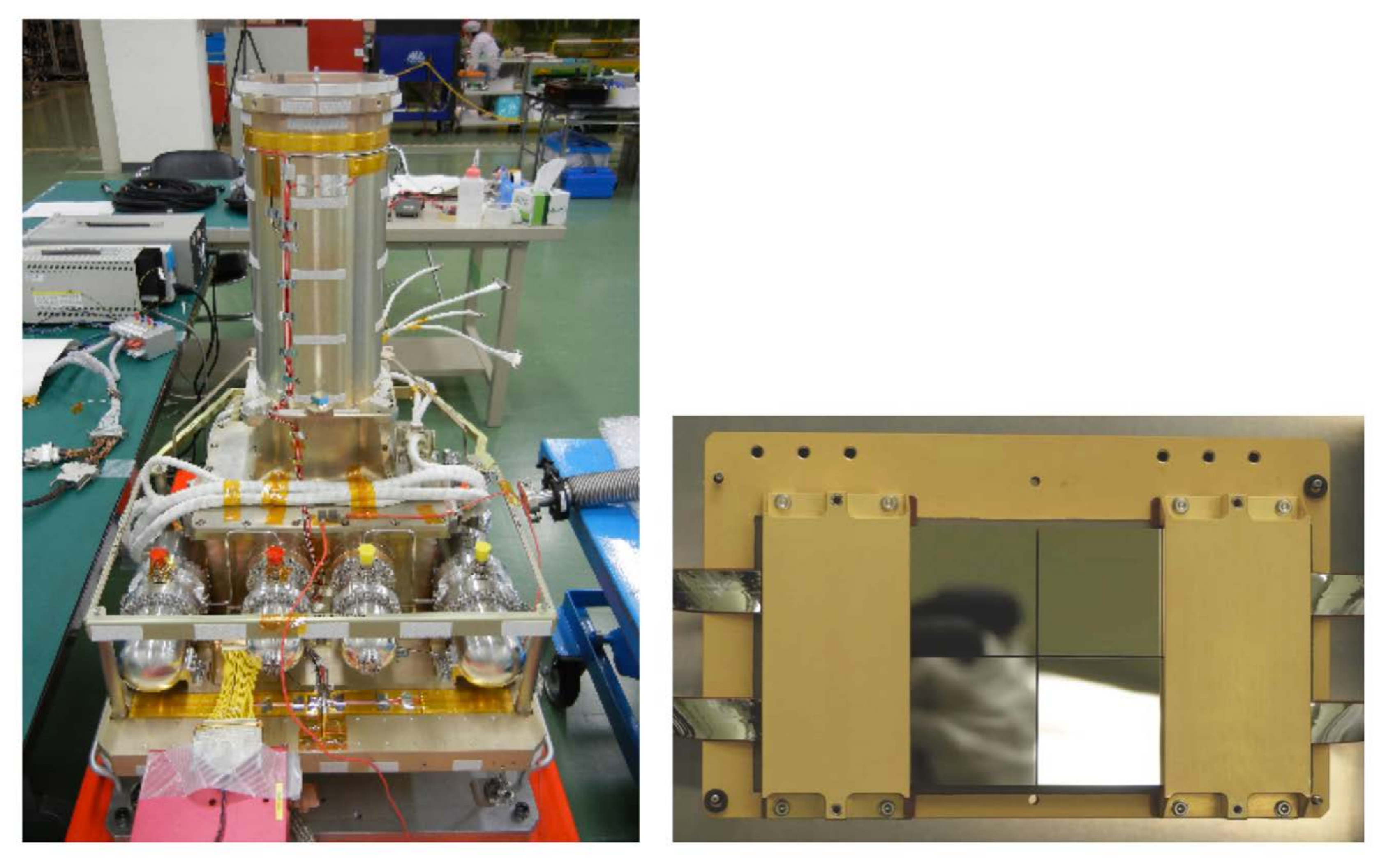}}

\caption{ (left) Photograph of  the SXI-S system, showing an overview (left).  The CCD detectors are mounted in the sensor body, underneath the hood, which is aligned with the telescope optical axis.  The hood also contains the contamination blocking filter.  (right) Photograph of the CCD detectors (flight model).   }
\label{SXI-OUTLOOK}
\end{figure}

\subsection{Soft X-ray Imaging System}
X-ray sensitive silicon charge-coupled devices (CCDs) are   key detectors
for X-ray astronomy. The low background and high energy resolution
achieved with the XIS/Suzaku clearly show that the X-ray CCD can also
play a very important role in the ASTRO-H mission. The  soft X-ray imaging system 
will consist of an imaging mirror, the Soft X-ray Telescope (SXT-I), and a CCD camera, the Soft X-ray Imager (SXI), as well as the cooling system\cite{Ref:Tsuru,Ref:Tsunemi,Ref:Tsunemi2010,Ref:Tsunemi2012,Ref:Hayashida2014}. 
Fig.~\ref{SXI-OUTLOOK}  (right) shows  a photograph of the SXI detector.

\begin{figure}[htb]
\centerline{\includegraphics[height=6cm]{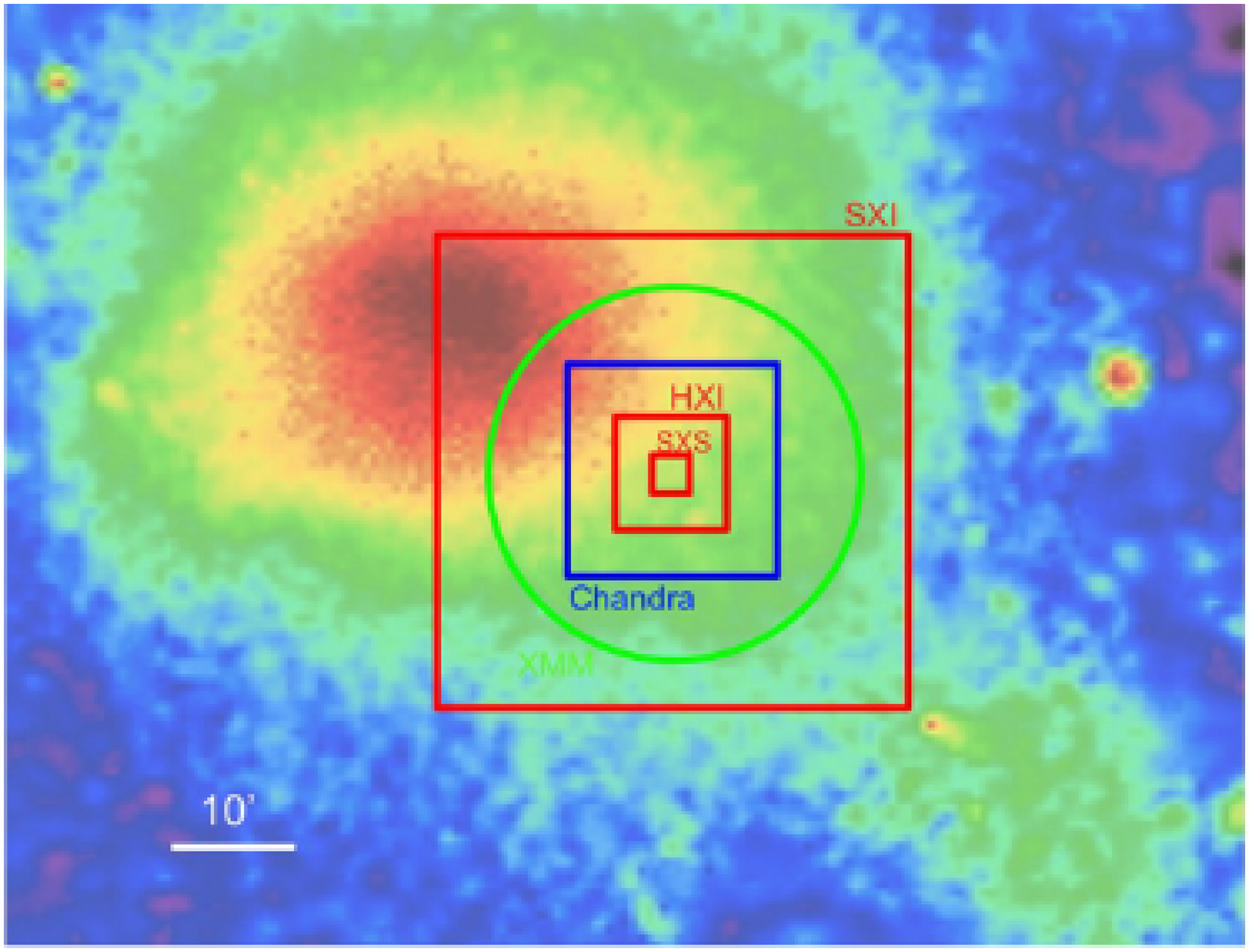}
\includegraphics[height=6cm,clip]{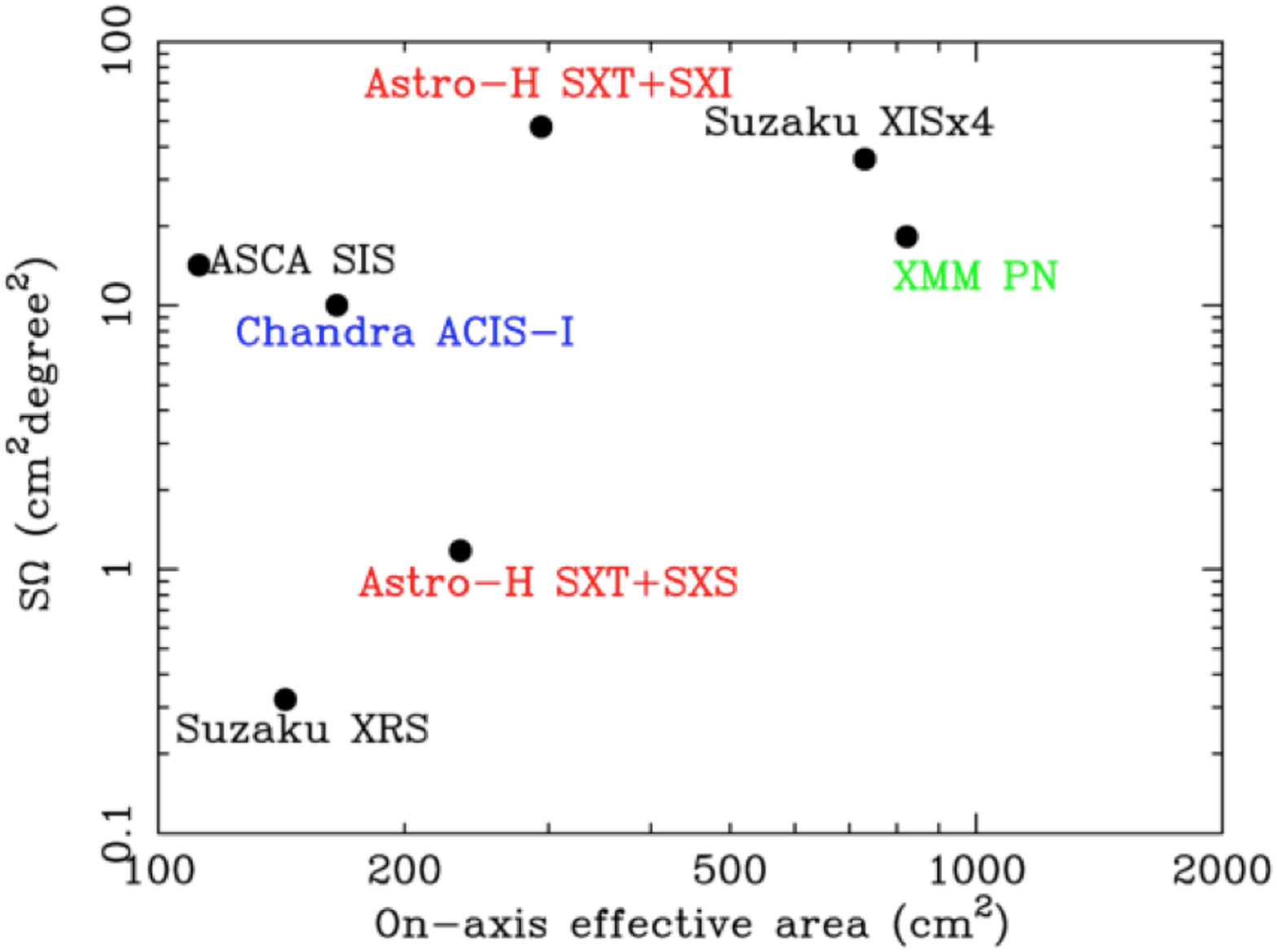}}
\caption{(left) Fields of view of the ASTRO-H instruments, SXS, SXI, HXI (red boxes). Chandra ACIS-I and XMM are also shown for comparison. The background image is the Coma cluster taken with ROSAT (credit: ROSAT/MPE/S. L. Snowden). (right) Grasp vs on-axis effective area at 7 keV of SXT-I+SXI, SXT-S+SXS.  Suzaku XIS, Chandra ACIS-I, XMM PN are also shown for comparison.
}
\label{Fig:SXIFOV}
\end{figure}

 In order to cover the soft X-ray band below 12~keV,
the SXI will use next generation Hamamatsu CCD chips with
a thick depletion layer of 200 $\mu$m, low noise, and almost no cosmetic defects. 
The  quantum efficiency 
is better than that  achieved by the Suzaku XIS over the entire 0.4--12 keV bandpass\@.  
The SXI-S contains four backside-illuminated (BI) charge-coupled devices (CCDs) with integrated frame storage.  The imaging area of each CCD is 3 cm on a side, spanning 1280 physical pixels each 24 $\mu$m in size.  The CCDs are arranged in a 2$\times$2 array, with a small gap of ~ 100 $\mu$m (3.5 arc sec) between the chips. 
A mechanical cooler ensures a long operational life at 
$-$120~$^\circ$C\@. 
The SXI features a large FOV and covers a 38$\times$38~arcmin$^{2}$ region on the sky,
complementing the smaller FOV of the 
SXS calorimeter (Fig. \ref{Fig:SXIFOV}). 
 To avoid target sources falling in the gap between the CCDs, the telescope aim point is offset 4.3 arcmin vertically and horizontally from the array center, co-aligned with the SXS and HXI fields of view, which are significantly smaller.


 \subsection{Hard X-ray Imaging System}

In addition to the improvement
of sensitivity brought by hard X-ray optics, the HXI provides a true imaging capability which enable us to
study spatial distributions of hard X-ray emission.

\begin{figure}[htb]
\centerline{\includegraphics[width=12cm]{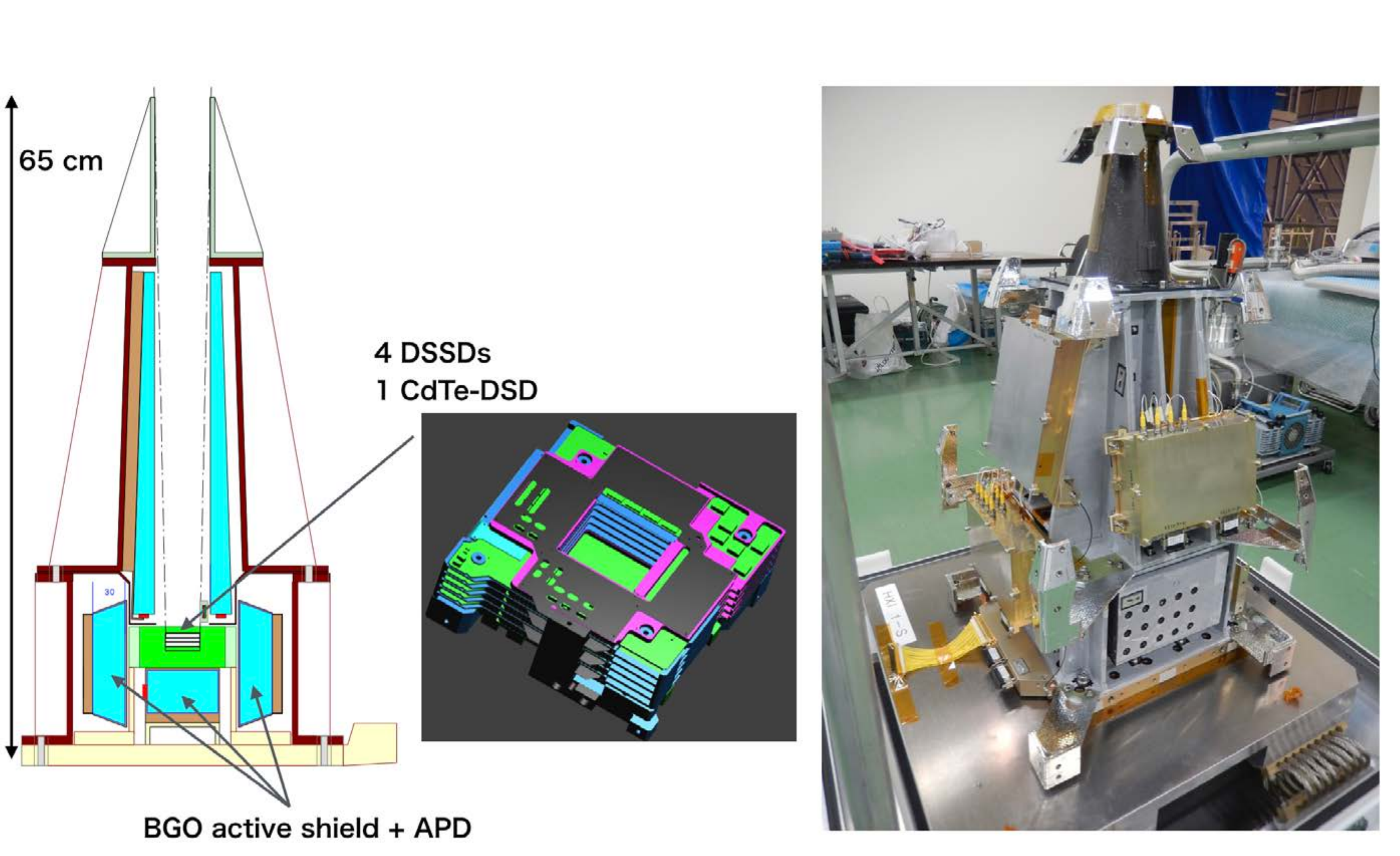}}
\caption{(left) Conceptual drawing of the Hard X-ray Imager. Each DSSD has a size of 3.2$\times$3.2~cm$^{2}$
and a thickness of 0.5~mm, resulting in 2~mm in total thickness, the same as that of the PIN detector 
of the HXD onboard Suzaku. A CdTe strip
detector has a size of 3.2$\times$3.2~cm$^{2}$ and a thickness of 0.75~mm. 
A stack of Si and CdTe double sided cross-strip detectors is
mounted in a well-type BGO shield. (right) Photograph  of  HXI-1 (flight model).
}
\label{Fig:HXI}
\end{figure}

The sensor part of the HXI consists of four layers of 0.5~mm thick Double-sided Silicon Strip
Detectors (DSSD) and one layer of 0.75~mm thick CdTe imaging 
detector (Fig.~\ref{Fig:HXI})\cite{Ref:Takahashi_SPIE1,Ref:Nakazawa,Ref:Kokubun,Ref:Kokubun2010,Ref:Kokubun2012,Ref:Sato2014}.  
In this configuration, soft X-ray photons below  $\sim$ 20 keV are 
absorbed in the Si part (DSSD), while hard X-ray photons above  $\sim$ 20 keV go through the
Si part and are detected by a newly developed CdTe double-sided cross-strip detector. 
The E$<$ 20 keV spectrum, obtained with  the  DSSD Si detector, has a much lower background due to the absence of activation in heavy material, such as Cd and Te. The DSSDs cover the energy below 30~keV while the 
CdTe strip detector covers the 20--80~keV band.
In addition to the increase in efficiency, the stack configuration and
individual readouts provide information on the interaction depth. This
depth information is very useful to reduce the background in space
applications, because we can expect that low energy X-rays interact in
the upper layers and, therefore, it is possible to reject the low energy events
detected in lower layers. Moreover, since the background rate scales
with the detector volume, low energy events collected from the first few
layers in the stacked detector have a high signal to background ratio,
in comparison with events obtained from a monolithic detector with a
thickness equal to the sum of all layers.  

In the energy band above 10 keV, the number of photons from the source decreases and the detector background becomes the major limitation  to its sensitivity. Since a significant fraction of the background events originate from interactions of the cosmic-ray with the detector structure, a tight active shield to reject cosmic-ray induced events is critical. 
Fast timing
response of the silicon strip detector and CdTe strip detector allows us to
place the entire detector inside a very deep well of an active shield
made of BGO (Bi$_{4}$Ge$_{3}$O$_{12}$) scintillators. The signal from the BGO
shield is used to reject background events.

\subsection{Soft Gamma-ray Detector (SGD)}

 The SGD measures soft $\gamma$-rays via reconstruction of  Compton scattering in the Compton camera, covering  an energy range of $40-600$\,keV
 with a sensitivity at 300 keV: 10 times better than that of the Suzaku Hard Xray
Detector.
It outperforms previous soft-$\gamma$-ray instruments in background
rejection capability by adopting a new concept of a narrow-FOV Compton
telescope, combining  Compton cameras and active well-type shields\cite{Ref:Takahashi_Yokohama,Ref:Takahashi_SiCdTe}.

\begin{figure}
\centerline{\includegraphics[height=10.0cm,clip]{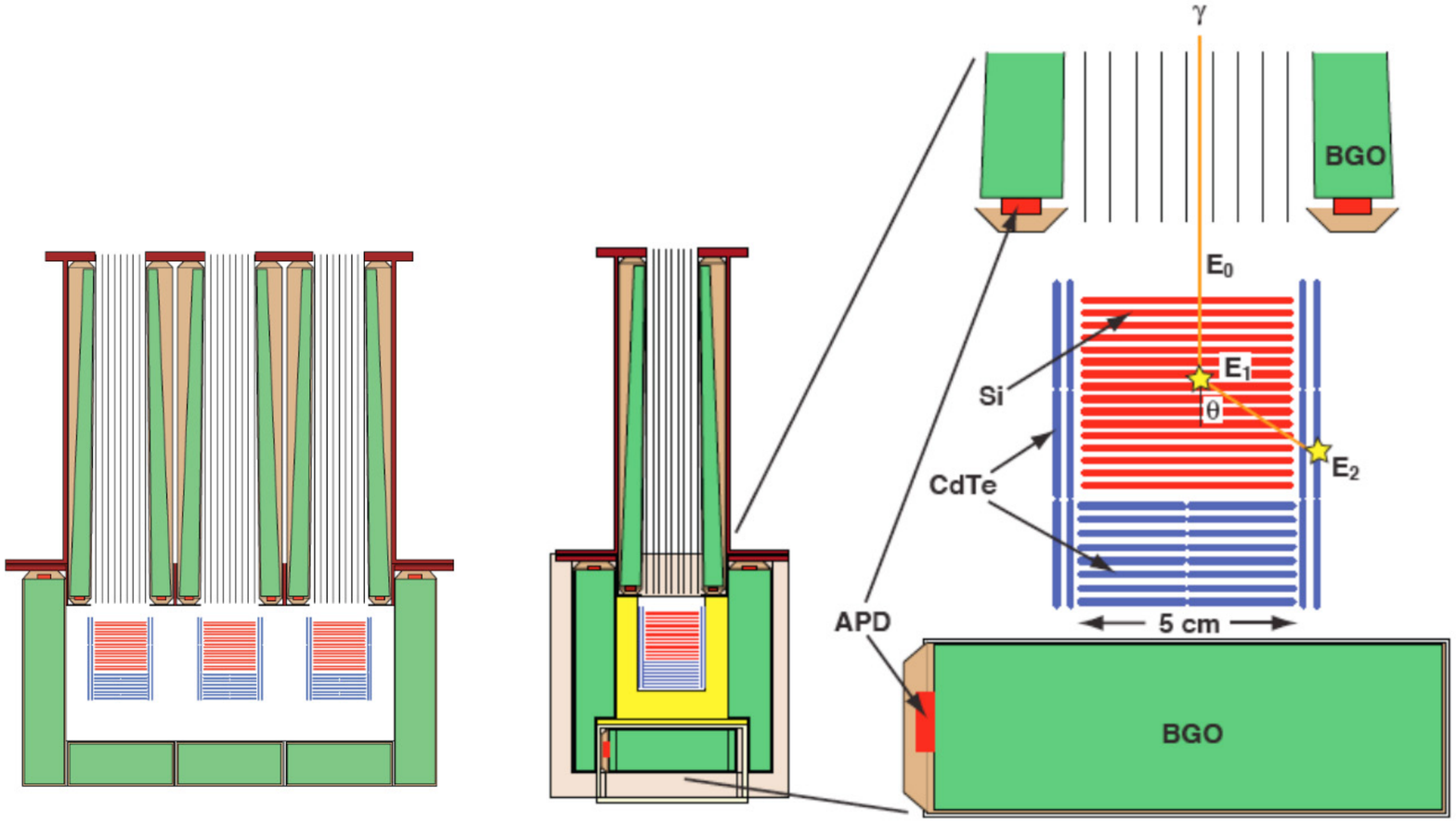}}
\caption{Conceptual drawing of an SGD Compton camera unit (green sections are the BGO anti-coincidence shields, red planes are the Si strip detectors in which the Compton scattering, occurs and the blue parts are the CdTe section in which the photons are absorbed).
In order to further restrict the FOV and to reduce  contamination from the cosmic X-ray background (CXB) for photons below 100 keV, 
a fine collimator is installed.}
\label{Fig:SGD_Concept}
\end{figure}

Figure \ref{Fig:SGD_Concept} shows the conceptual design of the Si/CdTe semiconductor Compton camera, together with two types of shield;
one is a BGO active shield and the other is a fine collimator made of PCuSn\cite{Ref:Takahashi_SiCdTe,Ref:Tajima2010,Ref:Watanabe2012,Ref:Fukazawa2014,Ref:Watanabe2014}
In the Si/CdTe Compton camera, events involving the incident gamma-ray being scattered 
in the Si detector and fully absorbed in the CdTe detectors are used for Compton imaging. 
The direction of the gamma-ray is 
calculated by solving the Compton kinematics with information concerning deposit energies and interaction positions recorded in the detectors.
In principle, each layer could act not only as a scattering part but also as an absorber part.
A very compact, high-angular resolution (fineness of image) camera is realized by fabricating 
semiconductor imaging elements made of Si and CdTe, 
which have excellent performance in position resolution, high-energy resolution, and high-temporal resolution.
As shown schematically in Fig.~\ref{Fig:SGD_Concept}, the detector
consists of 32 layers of 0.6~mm thick Si pad detectors 
and eight layers of CdTe pixellated detectors with a thickness of 0.75~mm.  The sides are 
also surrounded by two layers of CdTe pixel detectors.

\begin{figure}
\centerline{\includegraphics[height=6.0cm,clip]{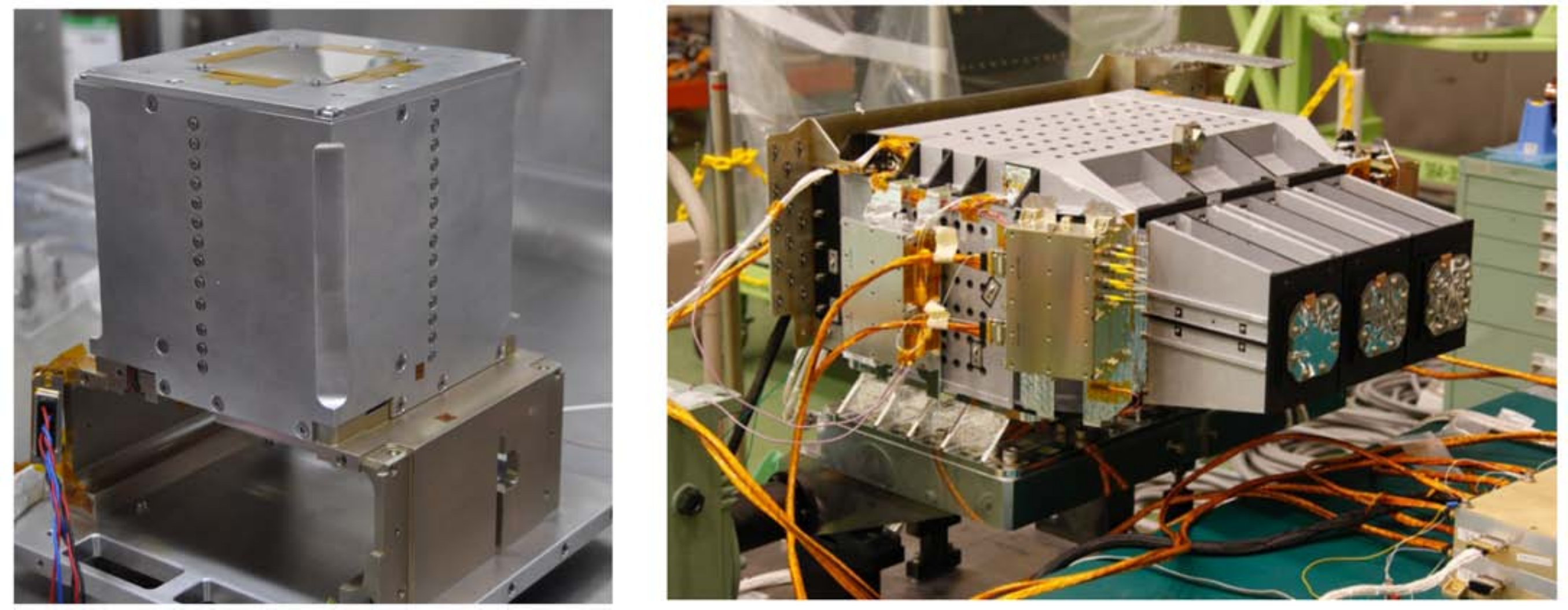}}
\caption{Photographs of  (left) a Si/CdTe Compton camera unit installed in a SGD and (right) SGD-1 (flight model). The SGD consists of two identical sets of an SGD-S (Sensor) which are called SGD-1 and SGD-2. The SGD-S is a detector body that includes a $3\times1$ array of identical Compton Camera Modules surrounded by BGO shield units and fine passive collimators. The two SGD-S are mounted on opposite sides of the spacecraft  side panels to balance the weight load since each has a large mass of 150 kg.  
}
\label{Fig:SGD_Photo}
\end{figure}

 The
camera is then mounted inside the bottom of a well-type active
shield (Fig.~\ref{Fig:SGD_Photo}).   The major advantage of employing a narrow FOV is that
the direction of incident $\gamma$-rays is constrained to be inside the FOV. If the Compton cone,
 which corresponds to the  direction of  incident gamma-rays, does not intercept the FOV, we can reject the event as background. Most of the background can be
rejected by requiring this condition\cite{Ref:Takahashi_Yokohama}.
The 
opening angle provided 
by the BGO shield is $\sim$10~degrees at 500~keV. 
An additional PCuSn collimator restricts the field of view of the
telescope to 30$^\prime$ for photons below 100 keV  to minimize the
flux due to the cosmic X-ray background in  the FOV.  
Since the  scattering angle of gamma-rays can be measured
via reconstruction of the Compton scattering in the Compton camera, the
SGD is capable of measuring polarization of celestial sources brighter
than a few $\times$ 1/100 of the Crab Nebula, if they are polarized by more than 10\%.
 This capability is expected to yield polarization measurements 
in several celestial objects, providing new insights into properties of soft gamma-ray
emission processes\cite{Ref:Tajima2010,Ref:Watanabe2014}.

In 2011, the technology behind the SGD was demonstrated  in measurements of the distribution of  Cs-137 in the environment
of Fukushima\cite{Ref:Compton_Fukushima}. In addition to showing that the SGD technology works as designed ``in the field'' the use of the Si/CdTe Compton Camera provided crucial information in understanding how the fallout from the 2011 nuclear accident was distributed.
%
%
%

\section{Expected Scientific Performance}

 ASTRO-H is expected to revolutize high energy science on all astronomical scales. These include: 1) the very compact scales around black holes, 2) high-temperature plasmas around stars at all stages of their lives, 3) the diffuse hot media  supernova remnants, 4) the interstellar media of galaxies, 5) clusters of galaxies and 6) the large scale diffuse inter galactic medium. Opening up new parameter space in i) energy range, ii) sensitivity, iii) spectral resolution and iv) polarimetry, ASTRO-H will allow detailed study of the dynamics, composition, morphology and evolution of matter across these cosmic scales.
In order to demonstrate the new science accessible with ASTRO-H, we are preparing a series of white papers by dedicated task forces. The categories of the task forces are listed in Table \ref{WPTF}.
In this section, new science topics enabled by ASTRO-H  are briefly described.
Further details are available in these white papers, which will become public,  soon.
 The high energy resolution of the non-dispersive Soft X-ray Spectrometer (SXS)   is unique in X-ray astronomy, since no other previously or currently operating spectrometers could  achieve comparable high energy resolution, high quantum efficiency, and spectroscopy for spatially extended sources at the same time. Imaging spectroscopy of extended sources can reveal line broadening and Doppler shifts due to turbulent or bulk velocities. 
This capability enables the determination of the level of turbulent pressure support in clusters, SNR ejecta dispersal patterns, the structure of AGN and starburst winds, and the spatially dependent abundance pattern in clusters and elliptical galaxies. The SXS can also measure the optical depths of resonance absorption lines, from which the degree and spatial extent of turbulence can be inferred. Additionally, the SXS can reveal the presence of relatively rare elements in SNRs and other sources through its high sensitivity to low equivalent width emission lines. The low SXS background ensures that the observations of almost all line-rich objects will be photon limited rather than background limited.

\begin{table}
\caption{Categories of ASTRO-H Science Task Forces}
\label{WPTF}
\begin{center}
\begin{tabular}{l}
\hline
Stars\\		
White dwarfs\\
Low-mass binaries	\\
High-mass binaries and magnetars	\\
Black hole spin and accretion	\\	 
Young SNRs		\\
Old SNRs and PWN		\\
Galactic center	\\	
ISM and galaxies		\\
Cluster-related sciences	\\
AGN reflection		\\
AGN winds		\\ 
New spectral features	\\	
Shocks and acceleration	\\	 
Broad-band spectra and polarization	\\	
High-z chemical evolution\\
\hline
\end{tabular}
\end{center}
\end{table}
 
\begin{figure}[htbp]
\includegraphics[width=16cm]{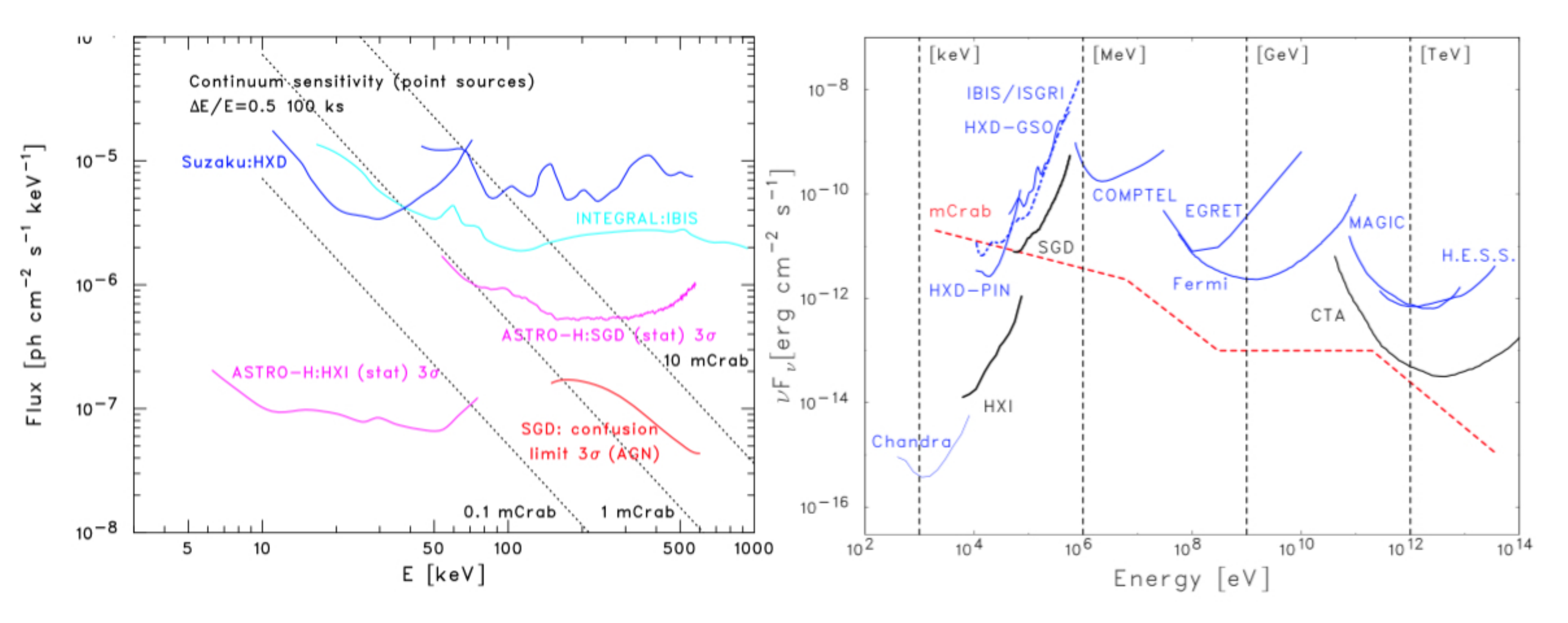}\caption{(left) The $3\sigma$ sensitivity curves for the HXI and SGD onboard ASTRO-H for an isolated point source. (100 ks exposures and $\Delta E/E = 0.5$) (right) Differential sensitivities of different X-ray and $\gamma$-ray instruments for an isolated point source\cite{Ref:Takahashi_CTA}. 
Lines for the \chandra/ACIS-S, the \suzaku/HXD (PIN and GSO), the \integral/IBIS (from the 2009 IBIS Observer's Manual), and the \astroh/HXI,SGD are the $3\sigma$ sensitivity curves for 100 ks exposures. A spectral bin with $\Delta E/E = 1$ is assumed for \chandra\ and $\Delta E/E = 0.5$ for the other instruments\cite{Ref:Takahashi_CTA}. }
\label{Fig:HXISGD}
\end{figure}

All studies of the total energy content of cosmic plasma (including that of non-thermal 
particles), aimed to draw a more complete picture of the
high energy universe, require observations by {\sl both} a spectrometer 
capable of measuring the bulk plasma velocities and/or turbulence with the 
resolution corresponding to the speed of a few $\times$ 100~km/s {\sl and}
an arc-min imaging system in the hard X-ray band, with  sensitivity two-orders of 
magnitude better than previous missions (see  Fig.~\ref{Fig:HXISGD}).
   The high energy resolution provided by SXS,  
 will make it possible to detect dozens of emission lines from highly ionized ions and measure, 
 for the first time, their line profiles with sufficient accuracy to study gas motions. 
 The Hard X-ray Imager (HXI) will extend the simultaneous spectral coverage to 
 energies well above 10 keV, which is critical for studying both thermal and non-thermal gas in clusters.
In bright, nearby galaxy clusters, such as the Perseus Cluster, ASTRO-H will determine the projected velocity 
$v_{bulk}$ (line centroid) and the line-of-sight velocity dispersion $\sigma_{v}$  (line width) as a
function of position, providing a measure of the bulk and small scale velocities of the plasmal\cite{Ref:WP_Cluster}.

 \begin{figure}
\centerline{\includegraphics[height=5.5cm,angle=0]{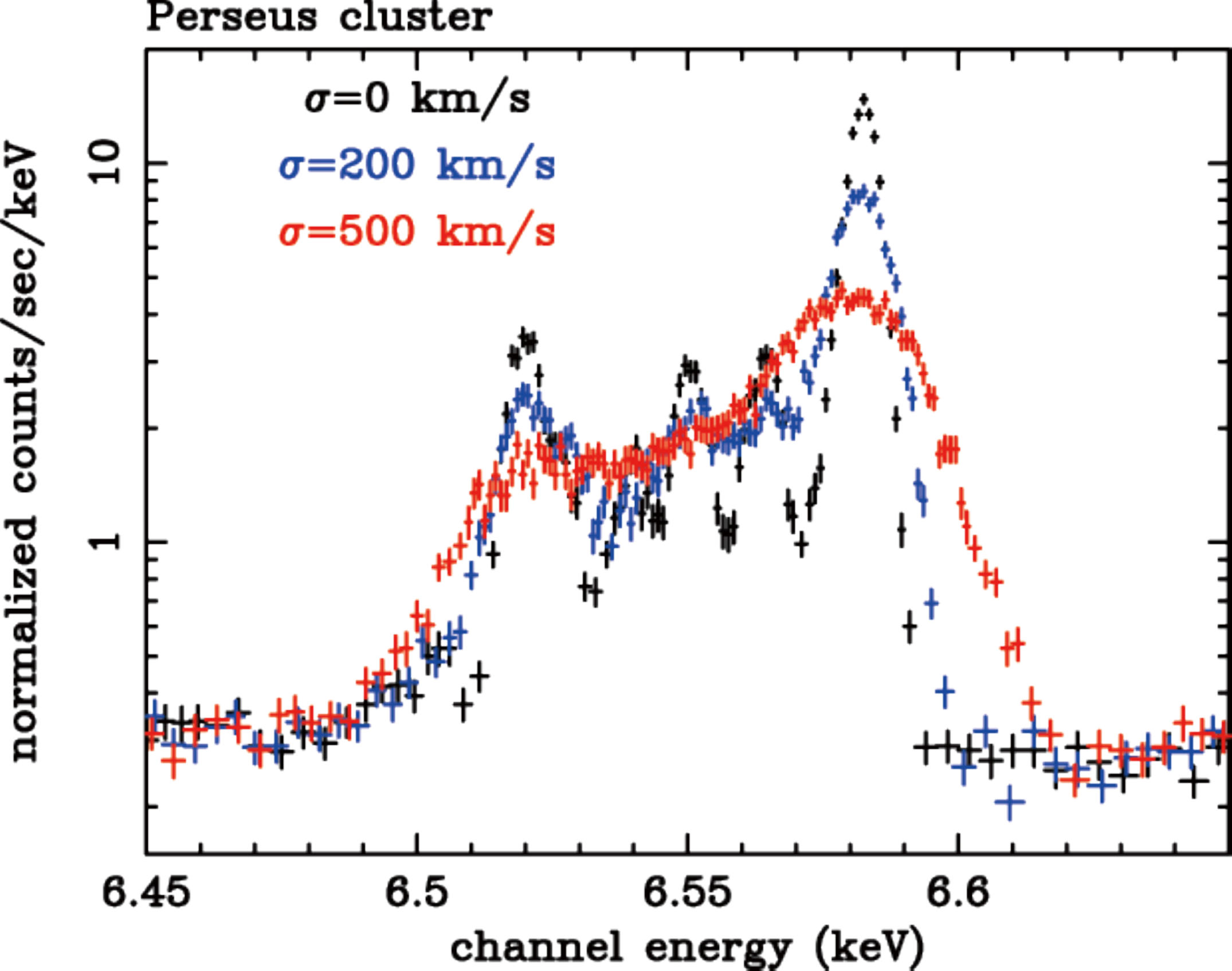}\hspace{5mm}
\includegraphics[height=5.5cm,angle=0]{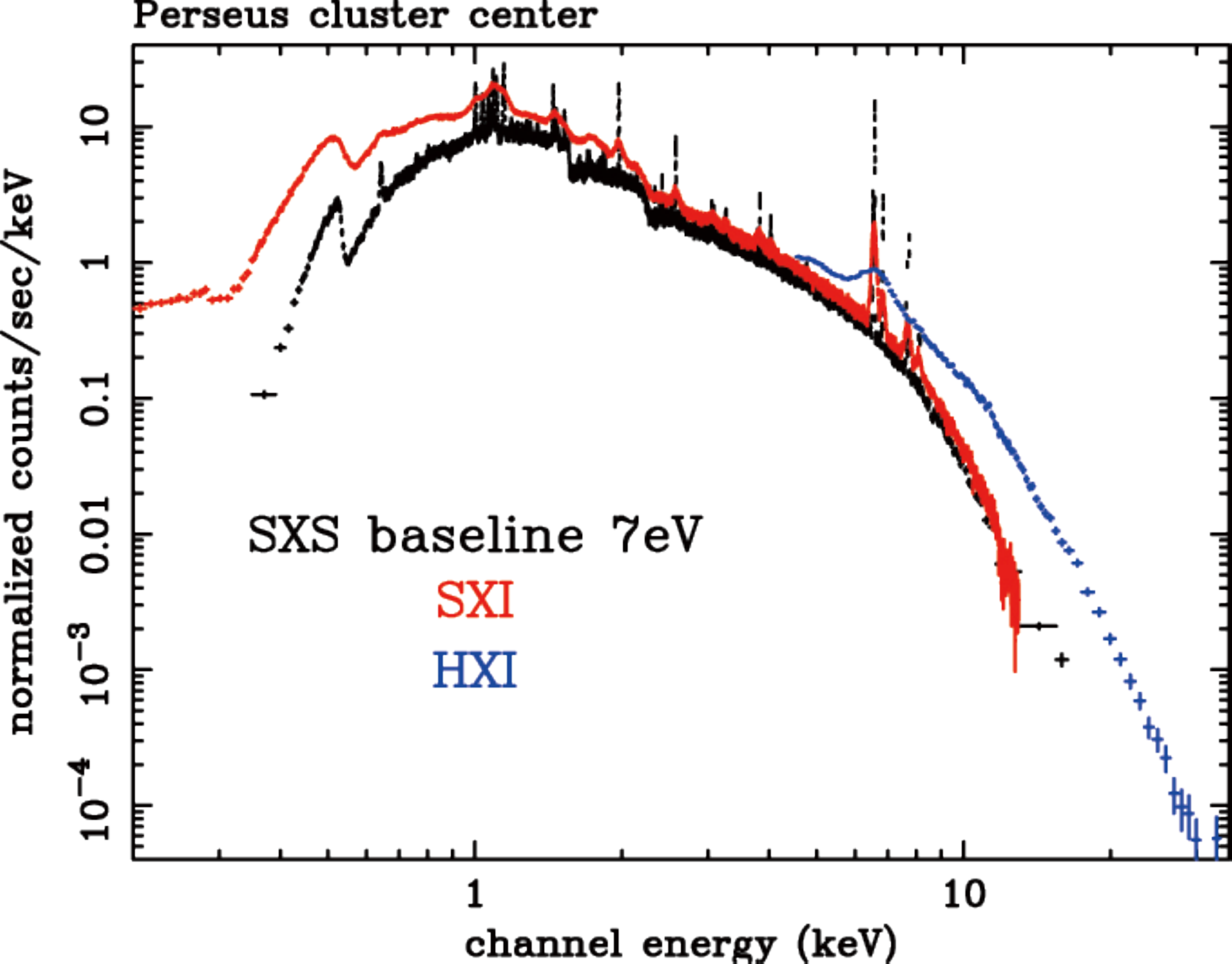}}
\caption{Simulated spectra for 100\,ks \astroh\ observations of Perseus Cluster. {\bf (left)} 
SXS spectra around the iron K line complex.  
Line profiles assuming $\sigma=0$, 200 and 500\,$\rm km\ s^{-1}$ turbulence. 
{\bf (right)} SXS (black), SXI (red), and HXI (blue) spectra 
for hot plasma with a mixture of three different temperatures of 0.6, 2.6 and 6.1\,keV ($r < 2'$)\cite{Ref:Takahashi2010}.}
\label{Fig:SXS2}\end{figure}

\begin{figure}[htbp]
\includegraphics[width=16cm]{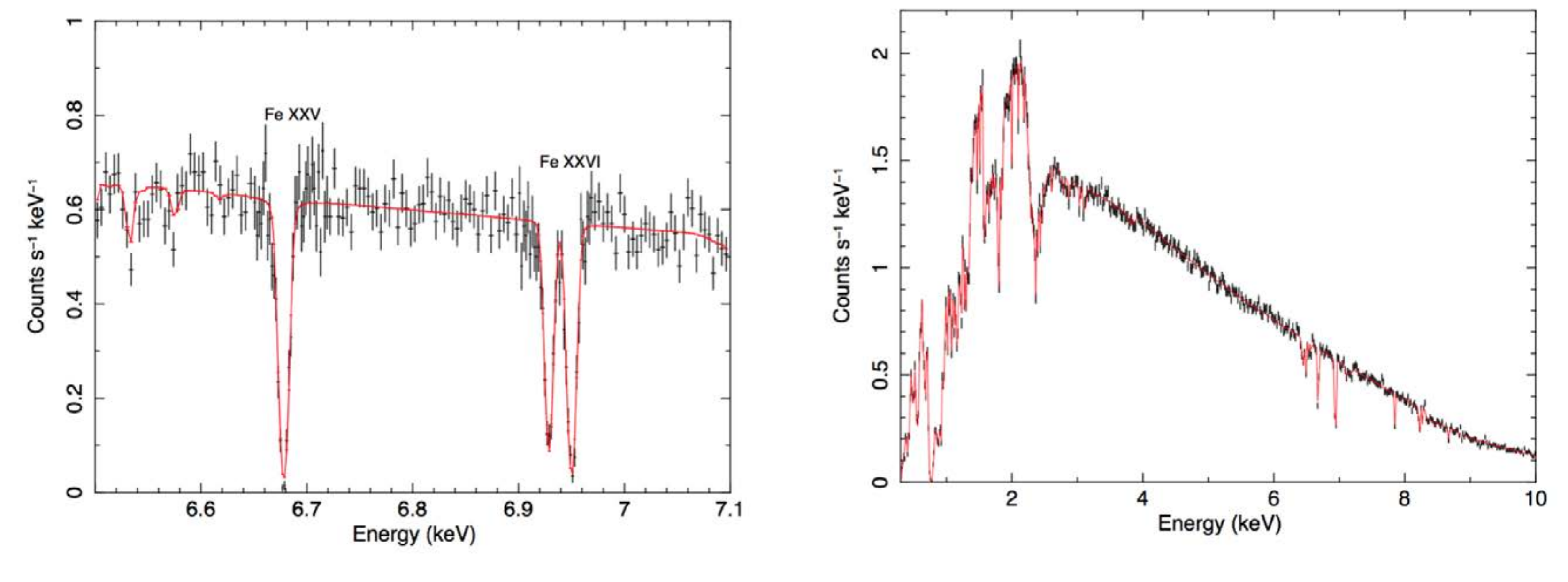}\caption{Simulated spectrum for NGC 4151 for 100 ks exposure time.
The same flux condition of the 2002 HETGS spectrum\cite{Ref:Kraemer} is used for simulation. }
\label{Fig:4151}
\end{figure}

XMM-Newton and Suzaku spectra of AGN frequently show time-variable absorption 
and emission features in the 5--10~keV band. If these features are due 
to Fe, they represent gas moving at very high velocities with both red-  
and blue-shifted components from material presumably near the event horizon. 
The sensitivity of the SXS in the Fe-K region extends the observed absorption measure distribution of the outflow up to the highest ionization states accessible. Due to the high-resolution and sensitivity it will also be able to give the definitive proof for the existence of ultra-fast outflows, and if so, characterize their physical properties in great detail. These ultra-fast outflows carry very large amounts of energy and momentum, and are of fundamental importance for feedback studies. 
ASTRO-H SXS observations of highly ionized outflows will measure the velocity, density and dynamics of the material, 
revolutionizing our understanding by determining the launch radius of these winds\cite{Ref:WP_AGNOutFlow}.
 A simulated spectrum for NGC 4151 is shown in Figure~\ref{Fig:4151}.

The imaging capabilities at high X-ray energies will open a new era in high spatial resolution studies of astrophysical sources of non-thermal emission above 10 keV, probed simultaneously with lower energy imaging spectroscopy.
This
will enable us to track  the evolution of active galaxies with accretion flows which are
heavily obscured, in order to accurately assess their contribution to
black hole growth over cosmological time.   
It will also uniquely allow
mapping of the spatial extent of the hard X-ray emission in diffuse sources, thus tracing the sites of cosmic ray acceleration in structures ranging in size from mega parsecs, such as clusters of galaxies, down to parsecs, such as young supernova remnants.  
Those studies will be complementary to the SXS measurements:  
observing the hard X-ray synchrotron emission will allow a study of the most energetic 
particles, thus revealing the details of particle acceleration mechanisms 
in supernova remnants, while the high resolution SXS data on the gas kinematics 
of the remnant will constrain the energy input into the accelerators.

As shown in Figure\,\ref{Fig:HXISGD}, the sensitivity to be achieved by \astroh\ (and similarly \nustar) 
is about 2 orders of magnitude better than previous collimated or coded mask instruments that have operated in this energy band.
 This will bring a breakthrough in our understanding of hard X-ray spectra of celestial sources in general. With this sensitivity, $30-50$\,\% of the hard X-ray Cosmic Background will be resolved. This will enable us to track the evolution of active galaxies with accretion flows that are heavily obscured, in order to accurately assess their contribution to the Cosmic X-ray Background (i.e., black hole growth)  over cosmic time.

There is a strong synergy between the hard X-ray imaging data and the high resolution ($\Delta E $ $\leqq$ 7~eV) soft X-ray spectrometer: the kinematics of the gas, probed by the width and energy of the emission lines, constrains the energetics of the system. 
The  kinematics of the gas provides information about the bulk motion; the energy of this motion is in turn
responsible for acceleration of particles to very high energies at
shocks, which is in turn manifested via non-thermal emission processes, best
studied via sensitive hard X-ray measurements.
All studies of the total energy content (including that of non-thermal 
particles), aimed to draw a more complete picture of the
high energy universe, require observations by {\sl both} a spectrometer 
capable of measuring the bulk plasma velocities and/or turbulence with the 
resolution corresponding to the speed of a few $\times$ 100~km/s {\sl and}
an arc-min imaging system in the hard X-ray band, with  sensitivity two orders of 
magnitude better than non-imaging missions. The power of ASTRO-H is that those gas dynamics can be probed both with micro calorimeters and the hard X-ray imaging instruments at the same time. Regarding the process of particle acceleration, the velocity field probed by the SXS data will tell us the conditions of the environment in which acceleration occurs, and the hard X-ray data will reveal how much acceleration is really taking place. Furthermore, the SGD data will tell us the maximum
energy of the accelerated particles.
 In this way, ASTRO-H will give us a new view of the non-thermal processes taking place in the universe.

 \section{Science Operation}
 
ASTRO-H will be launched into a circular orbit 
with altitude of 500--600~km, and inclination of 31~degrees.  Science 
operations will be similar to those of Suzaku, with pointed observation of each 
target until the integrated observing time is accumulated, and then slewing 
to the next target. A typical observation will require a few $\times$ 100~ksec integrated 
exposure time. All instruments  are co-aligned and 
will operate simultaneously. 

 ASTRO-H is in many ways similar to Suzaku in terms of orbit,
pointing, and tracking capabilities. After we launch the satellite, the current plan is to use the first three months for check-out and
start the PV phase with observations proprietary to the ASTRO-H team. Guest observing time will 
start from 10 months after the launch. About 75\% of the satellite time will be devoted to
 GO observations after the PV phase is completed. We are planning to implement key-project
type observations in conjunction with the GO observation time.

The telemetry from the satellite is downloaded and stored at the ground stations. The telemetry is distributed in real time to the operation control unit and quick look systems via SDTP protocol at the ground station. Then, the telemetry stored on each station is transferred to the SIRIUS database  in JAXA.  In the pre-pipeline process, the raw data are extracted from the SIRIUS database via SDTP protocol and stored into several FITS files, Raw Packet Telemetry (RPT), attitude file (ATT), orbit file (ORB), and command log (CMD).  The RPT contains all the information from the satellite. Since the RPT is just a dump of the space packets, the file is converted into the first FITS files (FFF), which contains the meaning of the attributes of the onboard instruments. 
 The data processing script run conversion of the FFF into calibrated event files, so--called second FITS file (SFF), corresponding to the Level 1 or unfiltered file, applies the data screening generating the Level 2 or cleaned files, and if appropriate extracts the Level 3 or data products such as spectra light curves and others. It also creates the so--called make filter file. All data files are in the FITS format and the software used in the pipeline is included in the standard ASTRO-H software package distributed to the science community. The data files archived include the Level 1, 2 and 3 corresponding to the unfiltered, cleaned and products files, the housekeeping data, orbit and attitude and the make filter file.

\section{Program Status}
\begin{figure}
\centerline{\includegraphics[width=11.0cm]{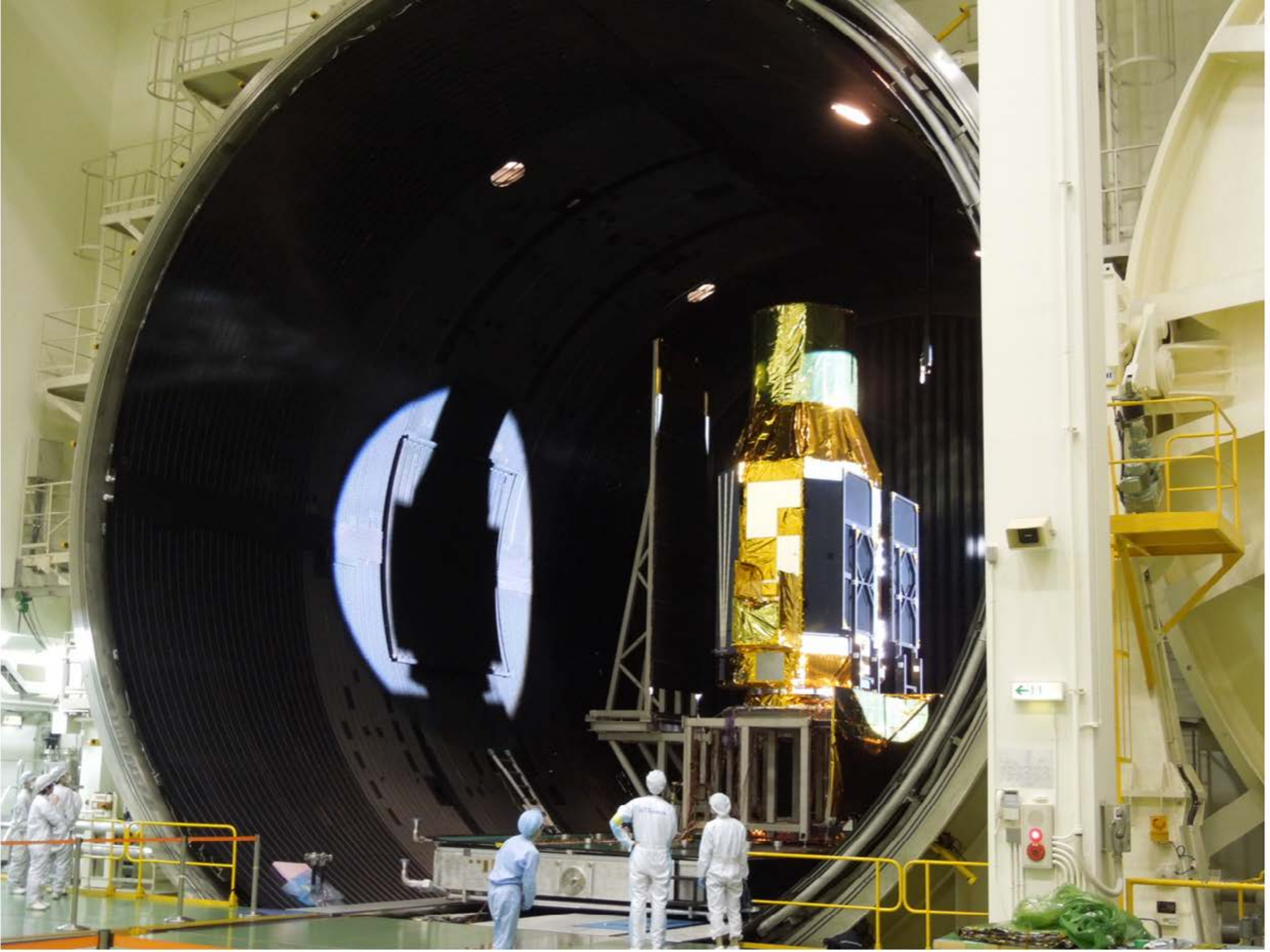}}
\caption{Photo taken at the thermal test (Aug 2012).}
\label{Fig:Photo1}
\end{figure}

\begin{figure}
\centerline{\includegraphics[width=11cm]{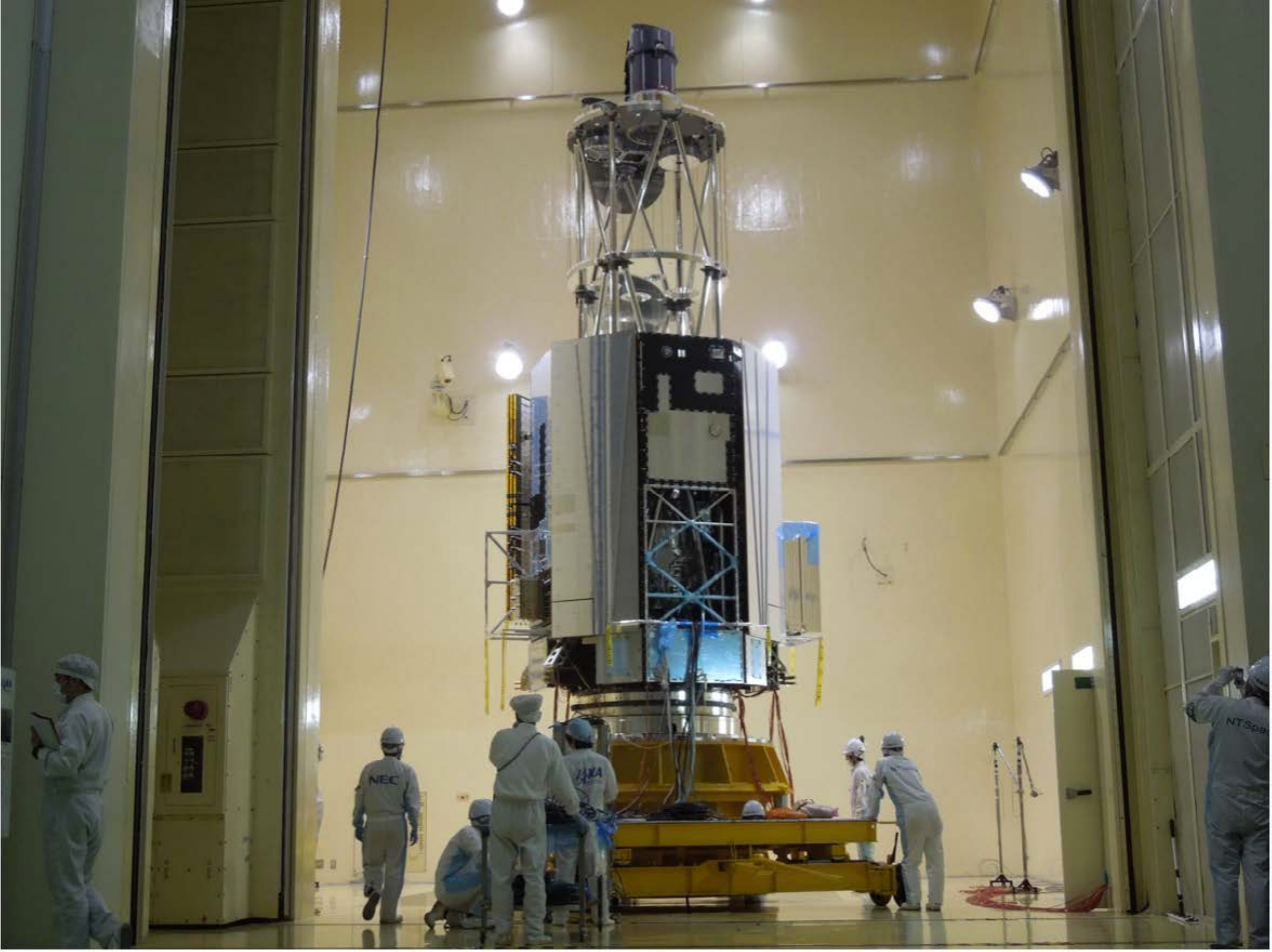}}
\caption{Photo taken at the acoustic test (May 2013).}
\label{Fig:Photo2}
\end{figure}
The ASTRO-H project officially started in JAXA in October 2008. The preliminary design review was held in May 2010.
After the detailed design phase (Phase C) was completed, the design of the satellite was reviewed and passed   the first critical design review (CDR1)
which was held in February 2012.  Since the thermal and mechanical design had to be verified well in advance, we started from manufacturing 
the  space craft structure, which  includes the side panels,
baseplates, and the FOB as preFM components. 
These components were assembled to form the thermal test model (TTM) and structural test model
(STM) of the satellite.
 A series of test campaigns were carried out  to ensure the validity of thermal design and mechanical design of the structure (Figs. \ref{Fig:Photo1} and \ref{Fig:Photo2})\cite{Ref:Iwata, Ref:Ishimura}.
In addition to usual thermal, acoustic, and vibration tests, we performed  a dedicated thermal deformation test
to verify the correctness of the design with respect to the alignment requirements for all co-aligned telescopes and instruments
on board ASTRO-H.
The thermal deformation during the ground test should be measured with an accuracy better  than 5 $\mu$m 
and 2 arc-seconds for  the size of the structure of about 10 meters.. To perform the thermal deformation test with such a high accuracy, a novel technique was developed and 
applied to ASTRO-H\cite{Ref:Ishimura}.

\begin{figure}
\centerline{\includegraphics[width=8.0cm]{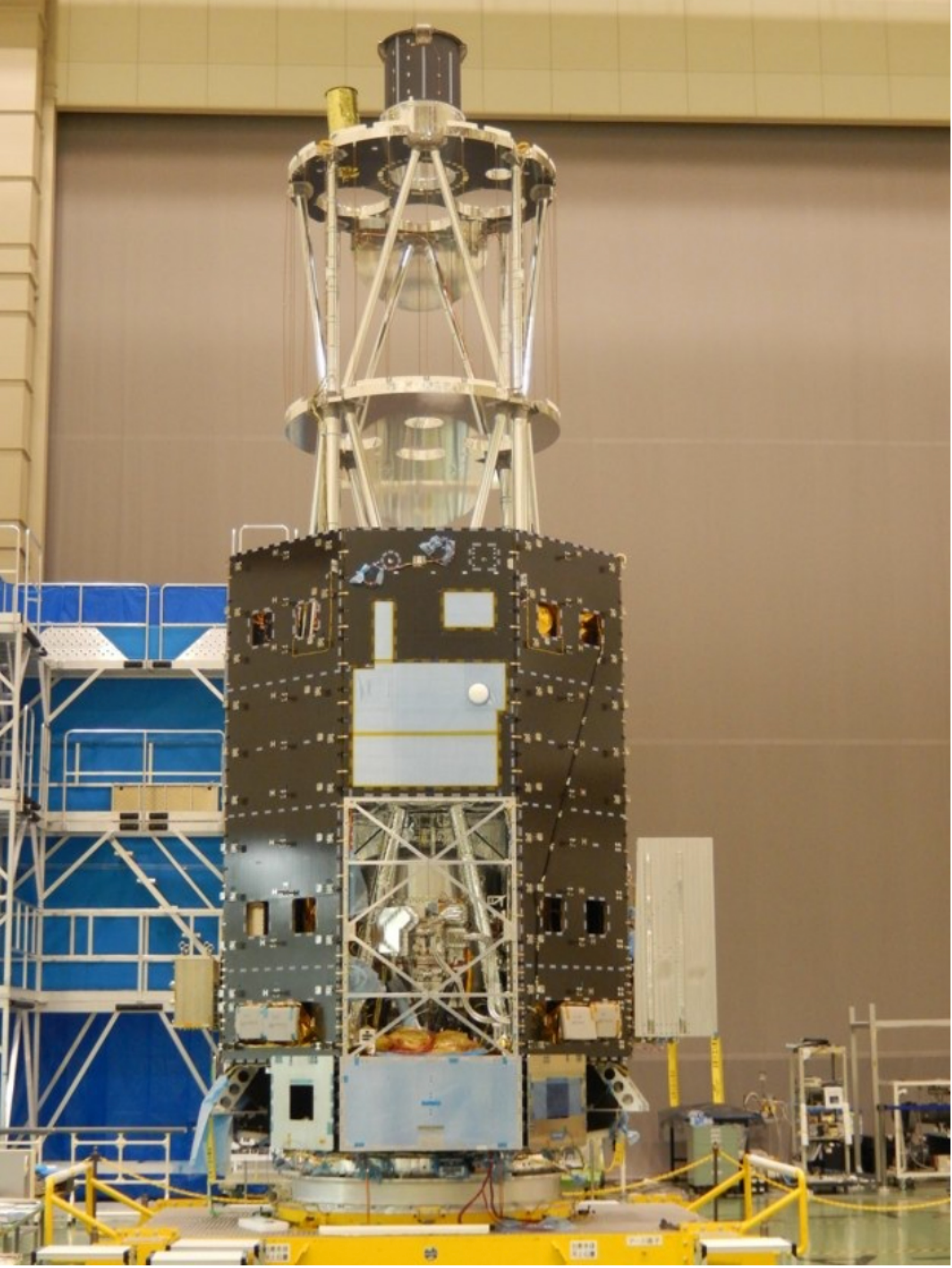}}
\caption{Photo taken at the first integration test (April 2014).}
\label{Fig:Photo3}
\end{figure}

By using most of FM components, the first  integration test campaign started in August 2013 (Fig. \ref{Fig:Photo3}).   The electrical and mechanical interfaces  between the satellite bus and the subsystem components, were verified, together with the  electrical power system of the satellite\cite{Ref:Shimada}, In June 2014, we successfully completed the test campaign. 
All  components mounted on the space craft structure are now being disassembled  for the final preparation. After we refurbish them, we will perform the final functional testing of each subsystem. For scientific instruments, the final calibration will also be carried out.  We are now planning to start the final integration and testing in November 2014.

\section{Summary}

ASTRO-H is scheduled to fly in 2015. 
Wide-band and high-resolution observations by the four instruments will provide exciting data sets for many science fields. 
The key properties of SXS onboard ASTRO-H are its high spectral resolution for both 
point and diffuse sources over a broad bandpass ($\leq$7~eV FWHM throughout 
the 0.3--12~keV band), high sensitivity (effective area of 160~cm$^2$ at 1~keV 
and 210~cm$^2$ at 7~keV), and low non-X-ray background 
(1.5$\times$10$^{-3}$~cts~s$^{-1}$keV$^{-1}$). These properties open up 
a full range of plasma diagnostics and kinematic studies of X-ray emitting 
gas for thousands of targets, both Galactic and extragalactic. The SXS improves 
upon and complements the current generation of X-ray missions, including 
Chandra, XMM-Newton, Suzaku, Swift and NuSTAR. 


ASTRO-H will also extend and enhance the completely new field of spatial studies of non-thermal emission across a broad range of energies extending well above 10 keV with hard X-ray telescopes and enable us to track the evolution of the dominant population 
of active galaxies with heavily obscured accretion flows.

\section*{Acknowledgments}

The authors are deeply grateful for on-going contributions provided by other members in the ASTRO-H team in Japan, the US, Europe and Canada. The team would like to acknowledge the valuable contribution of Henri Aarts
of SRON who unfortunately passed away last 
summer in 2013.

\end{document}